\documentclass[10pt,a4paper]{article}
\usepackage{amsfonts,amsmath,amssymb,cite,lscape}
\newcommand{\Real}{\mathop{\textrm{Re}}}
\newcommand{\sgn}{\mathop{\textrm{sgn}}}

\begin{document}
\title{Relativistic two-dimensional hydrogen-like atom \\
in a weak magnetic field}
\author{Rados{\l}aw Szmytkowski \\*[3ex]
Atomic and Optical Physics Division, \\
Department of Atomic, Molecular and Optical Physics, \\
Faculty of Applied Physics and Mathematics,
Gda{\'n}sk University of Technology, \\
ul.\ Gabriela Narutowicza 11/12, 80--233 Gda{\'n}sk, Poland \\
email: radoslaw.szmytkowski@pg.edu.pl}
\date{}
\maketitle
\begin{center}
\textbf{Published as: Ann.\ Phys.\ 401 (2019) 174--92} \\*[1ex]
\textbf{doi: 10.1016/j.aop.2018.12.007} \\*[5ex]
\end{center}
\begin{abstract} 
A two-dimensional (2D) hydrogen-like atom with a relativistic Dirac
electron, placed in a weak, static, uniform magnetic field
perpendicular to the atomic plane, is considered. Closed forms of the
first- and second-order Zeeman corrections to energy levels are
calculated analytically, within the framework of the
Rayleigh--Schr{\"o}dinger perturbation theory, for an arbitrary
electronic bound state. The second-order calculations are carried out
with the use of the Sturmian expansion of the two-dimensional
generalized radial Dirac--Coulomb Green function derived in the paper.
It is found that, in contrast to the case of the three-dimensional
atom [P.\ Stefa{\'n}ska, Phys.\ Rev.\ A 92 (2015) 032504], in two
spatial dimensions atomic magnetizabilities (magnetic
susceptibilities) are expressible in terms of elementary algebraic
functions of a nuclear charge and electron quantum numbers. The
problem considered here is related to the Coulomb impurity problem for
graphene in a weak magnetic field.
\mbox{} \\*[3ex]
\end{abstract}
%
%\newpage
%
\section{Introduction}
\label{I}
\setcounter{equation}{0}
Properties of model two-dimensional hydrogenic systems immersed in a
magnetic field have been investigated for several decades within the
frameworks of nonrelativistic \cite{Akim67,Gork68,Shin70,MacD86,%
Adam88,Edel89,Zhu90a,Zhu90b,Le93,Lehm95,Taut95,Vill98a,Robn03,Soyl06,%
Soyl07,Gade11,Hoan13a,Hoan13b,Esco14a,Esco15,Esco14b,Fera15,Flav15,%
Liu15,Arde16,Hoan16,Kova17a,Kova17b,Le17,Szmy18,Fern18} and
relativistic \cite{Vill98b,Khal99,Ho00,Must00,Vill01,Chia02,Vill03,%
Chia05,Akhm09,Rutk09,Posz10,Posz11} quantum mechanics. Besides of
being interesting from a purely theoretical point of view, results of
such studies are also important for understanding various aspects of
physics of low-dimensional semiconductors \cite{Akim67,Gork68,Shin70,%
Edel89,Zhu90a,Zhu90b,Lehm95,Vill98a,Soyl07,Hoan13b,Hoan16} and of
graphene \cite{Kand10,Gama11,Zhu12,Kim14,Souz14,Sun14,Zhu14,Kand15}.
The subject is still far from being exhausted, and further research in
this area, especially the one based on the use of analytical methods,
is certainly demanded.

The present paper meets that need. On the following pages, we shall
consider the planar hydrogen-like atom subjected to the action of a
static, uniform and weak magnetic field perpendicular to the atomic
plane. The main assumptions about the system are: (i) the interaction
potential between an electron and a nucleus, with the latter taken to
be point-like and spinless, is the three-dimensional one-over-distance
attractive Coulomb potential, (ii) the electron is relativistic in the
sense that its constrained planar dynamics is governed by the
two-dimensional Dirac equation. With these premises, within the
framework of the Rayleigh--Schr{\"o}dinger perturbation theory, we
shall derive closed-form analytical expressions for the first- and
second-order Zeeman corrections to an arbitrary atomic fine-structure
energy level. The reader will see that while the first-order
calculations, presented in Section \ref{III}, are straightforward and
require the knowledge of unperturbed planar Dirac--Coulomb
eigenfunctions only, the second-order analysis appears to be quite
challenging. As the standard sum-over-eigenstates formula for
$E^{(2)}$ is of no practical use in the present context (since the
energy spectrum of the Dirac--Coulomb Hamiltonian is mixed and in
addition to discrete eigenvalues it contains two scattering continua
as well), in Section \ref{IV} we shall exploit an alternative
representation of $E^{(2)}$ involving the radial Dirac--Coulomb
Sturmian functions. This will lead us eventually to a relatively
simple analytical formula for a magnetizability (magnetic
susceptibility) of a relativistic two-dimensional hydrogen-like atom
in an arbitrary discrete energy eigenstate.
%
%\newpage
%
\section{Setting the problem}
\label{II}
\setcounter{equation}{0}
Consider a planar one-electron atom with a motionless, point-like and
spinless nucleus of electric charge $Ze$, embedded in a static uniform
magnetic field of induction $\boldsymbol{B}$ perpendicular to the
atomic plane. Stationary energy levels of the atomic electron in such
a system are eigenvalues of the Dirac equation
\begin{subequations}
\begin{equation}
\left\{c\boldsymbol{\alpha}\cdot[-\mathrm{i}\hbar\boldsymbol{\nabla}
+e\boldsymbol{A}(\boldsymbol{r})]+\beta mc^{2}
-\frac{Ze^{2}}{(4\pi\epsilon_{0})r}-E\right\}\Psi(\boldsymbol{r})=0
\qquad (\boldsymbol{r}\in\mathbb{R}^{2}),
\label{2.1a}
\end{equation}
which is to be solved subject to the constraints that the wave
function $\Psi(\boldsymbol{r})$ is single-valued and forced to satisfy
the boundary conditions
\begin{equation}
\sqrt{r}\,\Psi(\boldsymbol{r})\stackrel{r\to0}{\longrightarrow}0,
\qquad
\sqrt{r}\,\Psi(\boldsymbol{r})\stackrel{r\to\infty}{\longrightarrow}0.
\label{2.1b}
\end{equation}
\label{2.1}%
\end{subequations}
In Eq.\ (\ref{2.1a}), $\boldsymbol{\alpha}$ is the matrix vector
defined as
\begin{equation}
\boldsymbol{\alpha}
=\alpha_{1}\boldsymbol{n}_{x}+\alpha_{2}\boldsymbol{n}_{y}
\label{2.2}
\end{equation}
($\boldsymbol{n}_{x}$ and $\boldsymbol{n}_{y}$ are the unit vectors
along axes of a Cartesian $\{x,y\}$ coordinate system in the atomic
plane), with
\begin{equation}
\alpha_{1}
=\left(
\begin{array}{cc}
0 & \sigma_{1} \\
\sigma_{1} & 0
\end{array}
\right),
\qquad
\alpha_{2}
=\left(
\begin{array}{cc}
0 & \sigma_{2} \\
\sigma_{2} & 0
\end{array}
\right),
\label{2.3}
\end{equation}
where $\sigma_{1}$ and $\sigma_{2}$ are the Pauli matrices
\begin{equation}
\sigma_{1}
=\left(
\begin{array}{cc}
0 & 1 \\
1 & 0
\end{array}
\right),
\qquad
\sigma_{2}
=\left(
\begin{array}{cc}
0 & -\mathrm{i} \\
\mathrm{i} & 0
\end{array}
\right),
\label{2.4}
\end{equation}
while $\beta$ is a $4\times4$ matrix of the form
\begin{equation}
\beta
=\left(
\begin{array}{cc}
I & 0 \\
0 & -I
\end{array}
\right),
\label{2.5}
\end{equation}
where $I$ is the unit $2\times2$ matrix. In the symmetric gauge used
in this work, the vector potential $\boldsymbol{A}(\boldsymbol{r})$ is
taken to be
\begin{equation}
\boldsymbol{A}(\boldsymbol{r})
=\frac{1}{2}\boldsymbol{B}\times\boldsymbol{r}.
\label{2.6}
\end{equation}

The Dirac equation (\ref{2.1a}) is separable in the standard polar
coordinates $r,\varphi$ (we choose the polar axis along the unit
vector $\boldsymbol{n}_{x}$), in the sense that it possesses
particular solutions of the form
\begin{equation}
\Psi_{n\kappa m_{\kappa}}(r,\varphi)
=\frac{1}{\sqrt{r}}
\left(
\begin{array}{c}
P_{n\kappa m_{\kappa}}(r)
\Phi_{\kappa m_{\kappa}}(\varphi) \\*[1ex]
\mathrm{i}Q_{n\kappa m_{\kappa}}(r)
\Phi_{-\kappa m_{\kappa}}(\varphi)
\end{array}
\right),
\label{2.7}
\end{equation}
where
\begin{equation}
\Phi_{\kappa m_{\kappa}}(\varphi)
=\frac{1}{\sqrt{2\pi}}
\left(
\begin{array}{c}
\delta_{-\kappa,m_{\kappa}}\,
\mathrm{e}^{\mathrm{i}(m_{\kappa}-1/2)\varphi}
\\*[1ex]
\delta_{\kappa,m_{\kappa}}\,
\mathrm{e}^{\mathrm{i}(m_{\kappa}+1/2)\varphi}
\end{array}
\right)
\qquad
(\textrm{${\textstyle\kappa=\pm\frac{1}{2},\pm\frac{3}{2},
\pm\frac{5}{2},\ldots}$; $m_{\kappa}=\pm\kappa$})
\label{2.8}
\end{equation}
are the axial spinors introduced by Poszwa and Rutkowski \cite{Posz10}
(for a summary of properties of these spinor functions, see Appendix
\ref{A} at the end of the present paper; the reader is warned that the
quantum number $\kappa$ we use here has the opposite sign in relation
to the one that appeared in Refs.\ \cite{Posz10,Posz11,Posz14}). If we
insert Eq.\ (\ref{2.7}) into Eq.\ (\ref{2.1a}), and then exploit the
identities (\ref{A.14}), (\ref{A.11b}) and (\ref{A.18}), we find that
the radial spinor
\begin{equation}
\psi_{n\kappa m_{\kappa}}(r)
=\left(
\begin{array}{c}
P_{n\kappa m_{\kappa}}(r) \\*[1ex]
Q_{n\kappa m_{\kappa}}(r)
\end{array}
\right)
\label{2.9}
\end{equation}
solves the equation
\begin{subequations}
\begin{equation}
\big[H_{\kappa m_{\kappa}}(r)-E_{n\kappa m_{\kappa}}\big]
\psi_{n\kappa m_{\kappa}}(r)=0
\label{2.10a}
\end{equation}
subject to the boundary conditions
\begin{equation}
\psi_{n\kappa m_{\kappa}}(r)\stackrel{r\to0}{\longrightarrow}0,
\qquad
\psi_{n\kappa m_{\kappa}}(r)\stackrel{r\to\infty}{\longrightarrow}0,
\label{2.10b}
\end{equation}
\label{2.10}%
\end{subequations}
with the radial Hamiltonian
\begin{equation}
H_{\kappa m_{\kappa}}(r)
=\left(
\begin{array}{cc}
{\displaystyle mc^{2}-\frac{Ze^{2}}{(4\pi\epsilon_{0})r}} &
{\displaystyle-c\hbar\left(-\frac{\mathrm{d}}{\mathrm{d}r}
+\frac{\kappa}{r}\right)-\frac{1}{2}\frac{m_{\kappa}}{\kappa}ecBr} 
\\*[2ex]
{\displaystyle-c\hbar\left(\frac{\mathrm{d}}{\mathrm{d}r}
+\frac{\kappa}{r}\right)-\frac{1}{2}\frac{m_{\kappa}}{\kappa}ecBr} &
{\displaystyle-mc^{2}-\frac{Ze^{2}}{(4\pi\epsilon_{0})r}}
\end{array}
\right)
\label{2.11}
\end{equation}
and with $E_{n\kappa m_{\kappa}}$ being the energy eigenvalue. We
label the eigensolutions with three quantum numbers $n$, $\kappa$,
$m_{\kappa}$. The latter two have been defined in Eq.\ (\ref{2.8})
(cf.\ also Appendix \ref{A}; the reader should observe that, in
contrast to the counterpart three-dimensional problem, in the present
case the quantum number $\kappa$ is a half-integer), while the first
one --- the principal quantum number $n$, is defined to be
\begin{equation}
n=n_{r}+|\kappa|+{\textstyle\frac{1}{2}}.
\label{2.12}
\end{equation}
The radial quantum number $n_{r}$ appearing in Eq.\ (\ref{2.12}) is
defined so that the number of nodes of $P_{n\kappa m_{\kappa}}(r)$ in
the open interval $(0,\infty)$ is $n_{r}$ for
$\kappa\leqslant-\frac{1}{2}$ (in that case $n_{r}\in\mathbb{N}_{0}$)
and $n_{r}-1$ for $\kappa\geqslant\frac{1}{2}$ (in that case
$n_{r}\in\mathbb{N}_{+}$).

For $Z\neq0$ and $B\neq0$, no general method of obtaining analytical
solutions to the system (\ref{2.10}) is known, and consequently one is
relied on the use of approximations. If the external magnetic field is
weak, as it will be assumed from now on, one may exploit the
Rayleigh--Schr{\"o}dinger perturbation theory. To this end, we split
the radial Hamiltonian (\ref{2.11}) in the following manner:
\begin{equation}
H_{\kappa m_{\kappa}}(r)=H_{\kappa}^{(0)}(r)
+H_{\kappa m_{\kappa}}^{(1)}(r),
\label{2.13}
\end{equation}
with the zeroth-order operator being the radial Dirac--Coulomb
Hamiltonian
\begin{equation}
H_{\kappa}^{(0)}(r)
=\left(
\begin{array}{cc}
{\displaystyle mc^{2}-\frac{Ze^{2}}{(4\pi\epsilon_{0})r}} &
{\displaystyle-c\hbar\left(-\frac{\mathrm{d}}{\mathrm{d}r}
+\frac{\kappa}{r}\right)} \\*[2ex]
{\displaystyle-c\hbar\left(\frac{\mathrm{d}}{\mathrm{d}r}
+\frac{\kappa}{r}\right)} &
{\displaystyle-mc^{2}-\frac{Ze^{2}}{(4\pi\epsilon_{0})r}}
\end{array}
\right)
\label{2.14}
\end{equation}
and with the perturbing operator being
\begin{equation}
H_{\kappa m_{\kappa}}^{(1)}(r)
=-\frac{1}{2}\frac{m_{\kappa}}{\kappa}ecBr
\left(
\begin{array}{cc}
0 & 1 \\
1 & 0
\end{array}
\right).
\label{2.15}
\end{equation}
In concordance with the partition (\ref{2.13}), we shall be seeking
solutions to the eigensystem (\ref{2.10}) in the form of the
Rayleigh--Schr{\"o}dinger series
\begin{subequations}
\begin{equation}
E_{n\kappa m_{\kappa}}=E_{n\kappa}^{(0)}
+E_{n\kappa m_{\kappa}}^{(1)}+E_{n\kappa m_{\kappa}}^{(2)}+\cdots
\label{2.16a}
\end{equation}
and
\begin{equation}
\psi_{n\kappa m_{\kappa}}(r)=\psi_{n\kappa}^{(0)}(r)
+\psi_{n\kappa m_{\kappa}}^{(1)}(r)
+\psi_{n\kappa m_{\kappa}}^{(2)}(r)+\cdots,
\label{2.16b}
\end{equation}
\label{2.16}%
\end{subequations}
where
\begin{equation}
\psi_{\dots}^{(k)}(r)
=\left(
\begin{array}{c}
P_{\dots}^{(k)}(r) \\*[1ex]
Q_{\dots}^{(k)}(r)
\end{array}
\right);
\label{2.17}
\end{equation}
the superscripts indicate orders of individual terms with respect to
the magnetic induction strength $B$.

The zeroth-order terms in the series (\ref{2.16a}) and (\ref{2.16b})
are solutions to the radial bound-state Dirac--Coulomb problem
\begin{subequations}
\begin{equation}
\big[H_{\kappa}^{(0)}(r)-E_{n\kappa}^{(0)}\big]
\psi_{n\kappa}^{(0)}(r)=0,
\label{2.18a}
\end{equation}
\begin{equation}
\psi_{n\kappa}^{(0)}(r)\stackrel{r\to0}{\longrightarrow}0,
\qquad
\psi_{n\kappa}^{(0)}(r)\stackrel{r\to\infty}{\longrightarrow}0.
\label{2.18b}
\end{equation}
\label{2.18}%
\end{subequations}
Solving the system (\ref{2.18}) as in the three-dimensional case,
bound-state energy levels of the electron in an isolated planar atom
are found to be
\begin{equation}
E_{n\kappa}^{(0)}=mc^{2}\frac{n_{r}+\gamma_{\kappa}}{N_{n_{r}\kappa}}
=\frac{mc^{2}}{\sqrt{\displaystyle1+\frac{(\alpha Z)^{2}}
{(n_{r}+\gamma_{\kappa})^{2}}}},
\label{2.19}
\end{equation}
where
\begin{equation}
N_{n_{r}\kappa}=\sqrt{n_{r}^{2}+2n_{r}\gamma_{\kappa}+\kappa^{2}}
\label{2.20}
\end{equation}
and
\begin{equation}
\gamma_{\kappa}=\sqrt{\kappa^{2}-(\alpha Z)^{2}},
\label{2.21}
\end{equation}
with $\alpha=e^{2}/(4\pi\epsilon_{0})c\hbar$ being the Sommerfeld
fine-structure constant. To ensure that $\gamma_{\kappa}$ is real for
all admitted values of $\kappa$, we impose the constraint
\begin{equation}
Z<\frac{1}{2}\alpha^{-1}.
\label{2.22}
\end{equation}
The corresponding radial wave functions, orthonormal in the sense of
\begin{equation}
\int_{0}^{\infty}\mathrm{d}r\:\psi_{n\kappa}^{(0)\textrm{T}}(r)
\psi_{n'\kappa}^{(0)}(r)=\delta_{nn'}
\label{2.23}
\end{equation}
(the superscript T denotes the matrix transpose), may be shown to have
the components
\begin{subequations}
\begin{eqnarray}
P_{n\kappa}^{(0)}(r) 
&=& \sqrt{\frac{Z\big(1+\epsilon_{n\kappa}^{(0)}\big)n_{r}!
(n_{r}+2\gamma_{\kappa})}
{2a_{0}N_{n_{r}\kappa}^{2}(N_{n_{r}\kappa}-\kappa)
\Gamma(n_{r}+2\gamma_{\kappa})}}
\left(\frac{2Zr}{N_{n_{r}\kappa}a_{0}}\right)^{\gamma_{\kappa}}
\exp\left(-\frac{Zr}{N_{n_{r}\kappa}a_{0}}\right)
\nonumber \\
&& \times\left[L_{n_{r}-1}^{(2\gamma_{\kappa})}
\left(\frac{2Zr}{N_{n_{r}\kappa}a_{0}}\right)
-\frac{N_{n_{r}\kappa}-\kappa}{n_{r}+2\gamma_{\kappa}}
L_{n_{r}}^{(2\gamma_{\kappa})}
\left(\frac{2Zr}{N_{n_{r}\kappa}a_{0}}\right)\right]
\label{2.24a}
\end{eqnarray}
and
\begin{eqnarray}
Q_{n\kappa}^{(0)}(r) 
&=& -\,\sqrt{\frac{Z\big(1-\epsilon_{n\kappa}^{(0)}\big)n_{r}!
(n_{r}+2\gamma_{\kappa})}
{2a_{0}N_{n_{r}\kappa}^{2}(N_{n_{r}\kappa}-\kappa)
\Gamma(n_{r}+2\gamma_{\kappa})}}
\left(\frac{2Zr}{N_{n_{r}\kappa}a_{0}}\right)^{\gamma_{\kappa}}
\exp\left(-\frac{Zr}{N_{n_{r}\kappa}a_{0}}\right)
\nonumber \\
&& \times\left[L_{n_{r}-1}^{(2\gamma_{\kappa})}
\left(\frac{2Zr}{N_{n_{r}\kappa}a_{0}}\right)
+\frac{N_{n_{r}\kappa}-\kappa}{n_{r}+2\gamma_{\kappa}}
L_{n_{r}}^{(2\gamma_{\kappa})}
\left(\frac{2Zr}{N_{n_{r}\kappa}a_{0}}\right)\right],
\label{2.24b}
\end{eqnarray}
\label{2.24}%
\end{subequations}
where $L_{n}^{(\alpha)}(\rho)$ stands for the generalized Laguerre
polynomial \cite[Sec.\ 5.5]{Magn66}; we define
$L_{-1}^{(\alpha)}(\rho)\equiv0$. In the above equations, for brevity
we have denoted
\begin{equation}
\epsilon_{n\kappa}^{(0)}=\frac{E_{n\kappa}^{(0)}}{mc^{2}}
=\frac{n_{r}+\gamma_{\kappa}}{N_{n_{r}\kappa}},
\label{2.25}
\end{equation}
while $a_{0}=(4\pi\epsilon_{0})\hbar^{2}/me^{2}$ is the Bohr radius.
The reader should observe that for $\kappa\geqslant\frac{1}{2}$ and
$n_{r}=0$ (i.e., for $n=\kappa+\frac{1}{2}$) the expressions in the
square braces in both Eqs.\ (\ref{2.24a}) and (\ref{2.24b}) do vanish.
Consequently, there are no planar Dirac--Coulomb bound states in that
case.

Each of the energy levels (\ref{2.19}) associated with a given value
of $n$ is seen to be fourfold degenerate (twice with respect to the
sign of $\kappa$ and, after the latter is fixed, twice with respect to
the sign of $m_{\kappa}$), except for the one for which
$\kappa=-(n-\frac{1}{2})$; the latter is only doubly degenerate (with
respect to the sign of $m_{\kappa}$). The sum of degeneracies of all
levels corresponding to a particular value of $n$ is $2(2n-1)$.

It is possible to classify planar atomic states according to a
quasi-spectroscopic scheme, proposed by Poszwa and Rutkowski
\cite{Posz10} and resembling the one used for atoms in three
dimensions. Within the framework of that scheme, which we shall adopt
hereafter, an atomic state with given quantum numbers $n$ and $\kappa$
is labeled as $nl_{|\kappa|}$, where
\begin{equation}
l=\left|\kappa+{\textstyle\frac{1}{2}}\right|
\label{2.26}
\end{equation}
(Poszwa and Rutkowski \cite{Posz10} defined
$l=\left|\kappa-\frac{1}{2}\right|$, but we recall that their $\kappa$
had the opposite sign), with the usual letter designation
\begin{equation}
\textrm{$l=0$ $\rightarrow$ s},
\quad
\textrm{$l=1$ $\rightarrow$ p},
\quad
\textrm{$l=2$ $\rightarrow$ d},
\quad
\textrm{etc.}
\label{2.27}
\end{equation}
Examples of the use of that classification scheme are given in Table
\ref{T.1}.
\begin{center}
[Place for Table \ref{T.1}]
\end{center}
%
%%\newpage
%
\section{The first-order Zeeman corrections to the Dirac--Coulomb
energy levels}
\label{III}
\setcounter{equation}{0}
Since the radial zeroth-order wave functions $\psi_{n\kappa}^{(0)}(r)$
are normalized to unity [cf.\ Eq.\ (\ref{2.23})], the first-order
contribution $E_{n\kappa m_{\kappa}}^{(1)}$ to $E_{n\kappa
m_{\kappa}}$ is simply given by
\begin{equation}
E_{n\kappa m_{\kappa}}^{(1)}=\int_{0}^{\infty}\mathrm{d}r\:
\psi_{n\kappa}^{(0)\textrm{T}}(r)H_{\kappa m_{\kappa}}^{(1)}(r)
\psi_{n\kappa}^{(0)}(r).
\label{3.1}
\end{equation}
With the use of Eqs.\ (\ref{2.15}) and (\ref{2.17}), Eq.\ (\ref{3.1})
may be cast into the form
\begin{equation}
E^{(1)}_{n\kappa m_{\kappa}}
=-\frac{m_{\kappa}}{\kappa}ecB\int_{0}^{\infty}\mathrm{d}r\:
rP_{n\kappa}^{(0)}(r)Q_{n\kappa}^{(0)}(r).
\label{3.2}
\end{equation}
The radial integral in Eq.\ (\ref{3.2}) may be taken with the aid of
the identity
\begin{equation}
\int_{0}^{\infty}\mathrm{d}x\:x^{\alpha+1}\mathrm{e}^{-x}
\big[L_{n}^{(\alpha)}(x)\big]^{2}
=\frac{(\alpha+2n+1)\Gamma(\alpha+n+1)}{n!}
\qquad (\Real\alpha>-2),
\label{3.3}
\end{equation}
which results from the general formula (cf.\ Ref.\ \cite[Eqs.\ (E54),
(E56) and (E60)]{Szmy97})
\begin{eqnarray}
&& \int_{0}^{\infty}\mathrm{d}x\:x^{\gamma}\mathrm{e}^{-x}
L_{n}^{(\alpha)}(x)L_{n'}^{(\beta)}(x)
=(-)^{n+n'}\sum_{k=0}^{\min(n,n')}\frac{\Gamma(k+\gamma+1)}{k!}
\binom{\gamma-\alpha}{n-k}\binom{\gamma-\beta}{n'-k}
\nonumber \\
&& \hspace*{25em} (\Real\gamma>-1).
\label{3.4}
\end{eqnarray}
Thus, one has
\begin{equation}
\int_{0}^{\infty}\mathrm{d}r\:
rP_{n\kappa}^{(0)}(r)Q_{n\kappa}^{(0)}(r)
=\frac{1}{4}\alpha a_{0}
\left[1-\frac{2\kappa(n_{r}+\gamma_{\kappa})}{N_{n_{r}\kappa}}\right],
\label{3.5}
\end{equation}
and consequently $E_{n\kappa m_{\kappa}}^{(1)}$ is found to be
\begin{equation}
E_{n\kappa m_{\kappa}}^{(1)}=-\frac{m_{\kappa}}{4\kappa}
\left[1-\frac{2\kappa(n_{r}+\gamma_{\kappa})}{N_{n_{r}\kappa}}\right]
\frac{B}{B_{0}}\frac{e^{2}}{(4\pi\epsilon_{0})a_{0}}.
\label{3.6}
\end{equation}
Here
\begin{equation}
B_{0}=\frac{\hbar}{ea_{0}^{2}}
=\frac{m^{2}e^{3}}{(4\pi\epsilon_{0})^{2}\hbar^{3}}
\simeq2.35\times10^{5}~\textrm{T}
\label{3.7}
\end{equation}
is the atomic unit of magnetic induction. For states with $n_{r}=0$
(i.e., those with $\kappa=-n+\frac{1}{2}$), Eq.\ (\ref{3.6})
simplifies and gives
\begin{equation}
E_{n,-n+1/2,m_{-n+1/2}}^{(1)}
=\frac{m_{-n+1/2}}{4\left(n-\frac{1}{2}\right)}(2\gamma_{n-1/2}+1)
\frac{B}{B_{0}}\frac{e^{2}}{(4\pi\epsilon_{0})a_{0}}.
\label{3.8}
\end{equation}
%
%%\newpage
%
\section{The second-order Zeeman corrections to the Dirac--Coulomb
energy levels. Atomic magnetizabilities}
\label{IV}
\setcounter{equation}{0}
The Rayleigh--Schr{\"o}dinger perturbation theory gives the following
expression for the second-order correction to energy:
\begin{equation}
E_{n\kappa m_{\kappa}}^{(2)}=\int_{0}^{\infty}\mathrm{d}r\:
\psi_{n\kappa}^{(0)\textrm{T}}(r)H_{\kappa m_{\kappa}}^{(1)}(r)
\psi_{n\kappa m_{\kappa}}^{(1)}(r).
\label{4.1}
\end{equation}
Here $\psi_{n\kappa m_{\kappa}}^{(1)}(r)$ is the first-order
contribution to the radial spinor wave function. It solves the
inhomogeneous equation
\begin{subequations}
\begin{equation}
\big[H_{\kappa}^{(0)}(r)-E_{n\kappa}^{(0)}\big]
\psi_{n\kappa m_{\kappa}}^{(1)}(r)
=-\big[H_{\kappa m_{\kappa}}^{(1)}(r)
-E_{n\kappa m_{\kappa}}^{(1)}\big]\psi_{n\kappa}^{(0)}(r),
\label{4.2a}
\end{equation}
with $E_{n\kappa m_{\kappa}}^{(1)}$ determined in Section \ref{III},
subject to the boundary conditions
\begin{equation}
\psi_{n\kappa m_{\kappa}}^{(1)}(r)
\stackrel{r\to0}{\longrightarrow}0,
\qquad
\psi_{n\kappa m_{\kappa}}^{(1)}(r)
\stackrel{r\to\infty}{\longrightarrow}0
\label{4.2b}
\end{equation}
and subject to the further constraint
\begin{equation}
\int_{0}^{\infty}\mathrm{d}r\:\psi_{n\kappa}^{(0)\textrm{T}}(r)
\psi_{n\kappa m_{\kappa}}^{(1)}(r)=0.
\label{4.2c}
\end{equation}
\label{4.2}%
\end{subequations}
A formal solution to the system (\ref{4.2}) is
\begin{equation}
\psi_{n\kappa m_{\kappa}}^{(1)}(r)=-\int_{0}^{\infty}\mathrm{d}r'\:
\hat{G}_{n\kappa}^{(0)}(r,r')
\big[H_{\kappa m_{\kappa}}^{(1)}(r')-E_{n\kappa m_{\kappa}}^{(1)}\big]
\psi_{n\kappa}^{(0)}(r'),
\label{4.3}
\end{equation}
where $\hat{G}_{n\kappa}^{(0)}(r,r')$ is a generalized radial
Dirac--Coulomb Green function associated with the unperturbed Coulomb
energy level $E_{n\kappa}^{(0)}$. The function
$\hat{G}_{n\kappa}^{(0)}(r,r')$ is defined to be a solution to
the inhomogeneous Dirac--Coulomb equation
\begin{subequations}
\begin{equation}
\big[H_{\kappa}^{(0)}(r)-E_{n\kappa}^{(0)}\big]
\hat{G}_{n\kappa}^{(0)}(r,r')=\delta(r-r')I
-\psi_{n\kappa}^{(0)}(r)\psi_{n\kappa}^{(0)\textrm{T}}(r'),
\label{4.4a}
\end{equation}
subject to the boundary conditions
\begin{equation}
\hat{G}_{n\kappa}^{(0)}(r,r')
\stackrel{r\to0}{\longrightarrow}0,
\qquad
\hat{G}_{n\kappa}^{(0)}(r,r')
\stackrel{r\to\infty}{\longrightarrow}0,
\label{4.4b}
\end{equation}
together with the orthogonality constraint
\begin{equation}
\int_{0}^{\infty}\mathrm{d}r\:\psi_{n\kappa}^{(0)\textrm{T}}(r)
\hat{G}_{n\kappa}^{(0)}(r,r')=0.
\label{4.4c}
\end{equation}
\label{4.4}%
\end{subequations}
It is a $2\times2$ matrix-valued function and since the Dirac--Coulomb
operator is self-adjoint, it is symmetric in the sense of
\begin{equation}
\hat{G}_{n\kappa}^{(0)\textrm{T}}(r,r')
=\hat{G}_{n\kappa}^{(0)}(r',r).
\label{4.5}
\end{equation}
Application of Eq.\ (\ref{4.5}) to Eq.\ (\ref{4.4c}) implies the
orthogonality relation
\begin{equation}
\int_{0}^{\infty}\mathrm{d}r'\:\hat{G}_{n\kappa}^{(0)}(r,r')
\psi_{n\kappa}^{(0)}(r')=0,
\label{4.6}
\end{equation}
which simplifies Eq.\ (\ref{4.3}) to the form
\begin{equation}
\psi_{n\kappa m_{\kappa}}^{(1)}(r)=-\int_{0}^{\infty}\mathrm{d}r'\:
\hat{G}_{n\kappa}^{(0)}(r,r')
H_{\kappa m_{\kappa}}^{(1)}(r')\psi_{n\kappa}^{(0)}(r').
\label{4.7}
\end{equation}
Upon insertion of Eq.\ (\ref{4.7}) into Eq.\ (\ref{4.1}), we obtain
the following formula for the second-order energy correction
$E_{n\kappa m_{\kappa}}^{(2)}$:
\begin{equation}
E_{n\kappa m_{\kappa}}^{(2)}
=-\int_{0}^{\infty}\mathrm{d}r\int_{0}^{\infty}\mathrm{d}r'\:
\psi_{n\kappa}^{(0)\textrm{T}}(r)H_{\kappa m_{\kappa}}^{(1)}(r)
\hat{G}_{n\kappa}^{(0)}(r,r')H_{\kappa m_{\kappa}}^{(1)}(r')
\psi_{n\kappa}^{(0)}(r').
\label{4.8}
\end{equation}
Application of Eqs.\ (\ref{2.15}) and (\ref{2.17}) casts Eq.\
(\ref{4.8}) into the form
\begin{equation}
E_{n\kappa}^{(2)}=-\frac{1}{4}e^{2}c^{2}B^{2}
\int_{0}^{\infty}\mathrm{d}r\int_{0}^{\infty}\mathrm{d}r'
\left(
\begin{array}{cc}
Q_{n\kappa}^{(0)}(r) & P_{n\kappa}^{(0)}(r)
\end{array}
\right)
r\hat{{G}}_{n\kappa}^{(0)}(r,r')r'
\left(
\begin{array}{c}
Q_{n\kappa}^{(0)}(r') \\*[1ex]
P_{n\kappa}^{(0)}(r')
\end{array}
\right),
\label{4.9}
\end{equation}
where we have made use of the fact that the ratio $m_{\kappa}/\kappa$
is of unit modulus. Since the right-hand side of the above equation is
evidently independent of $m_{\kappa}$, the third subscript at
$E^{(2)}$ has been, and henceforth will be, dropped.

To evaluate the double integral in Eq.\ (\ref{4.9}), one has to insert
into the integrand some particular explicit representation of the
generalized Green function $\hat{{G}}_{n\kappa}^{(0)}(r,r')$.
The one we shall use here has a form of a series expansion in the
radial Dirac--Coulomb Sturmian basis. We shall construct that
expansion below, omitting most details since the procedure is very
much analogous to the one we have developed for three-dimensional
problems \cite{Szmy97}.

The discrete radial Dirac--Coulomb Sturmian functions for the problem
at hand are defined to be solutions to the differential eigensystem
\begin{subequations}
\begin{eqnarray}
&& \left(
\begin{array}{cc}
{\displaystyle mc^{2}-E
-\mu_{n_{r}^{\prime}\kappa}^{(0)}(E)
\frac{Ze^{2}}{(4\pi\epsilon_{0})r}} &
-{\displaystyle c\hbar\left(-\frac{\mathrm{d}}{\mathrm{d}r}
+\frac{\kappa}{r}\right)} \\*[2ex]
-{\displaystyle c\hbar\left(\frac{\mathrm{d}}{\mathrm{d}r}
+\frac{\kappa}{r}\right)} &
{\displaystyle-mc^{2}-E
-\mu_{n_{r}^{\prime}\kappa}^{(0)\,-1}(E)
\frac{Ze^{2}}{(4\pi\epsilon_{0})r}}
\end{array}
\right)
\left(
\begin{array}{c}
S_{n_{r}^{\prime}\kappa}^{(0)}(E,r) \\*[1ex]
T_{n_{r}^{\prime}\kappa}^{(0)}(E,r)
\end{array}
\right)
=0,
\nonumber \\
&&
\label{4.10a}
\end{eqnarray}
\begin{equation}
S_{n_{r}^{\prime}\kappa}^{(0)}(E,r)
\stackrel{r\to0}{\longrightarrow}0,
\qquad
T_{n_{r}^{\prime}\kappa}^{(0)}(E,r)
\stackrel{r\to0}{\longrightarrow}0,
\label{4.10b}
\end{equation}
\begin{equation}
S_{n_{r}^{\prime}\kappa}^{(0)}(E,r)
\stackrel{r\to\infty}{\longrightarrow}0,
\qquad
T_{n_{r}^{\prime}\kappa}^{(0)}(E,r)
\stackrel{r\to\infty}{\longrightarrow}0.
\label{4.10c}
\end{equation}
\label{4.10}%
\end{subequations}
Here $E$ is a fixed, real, energy-dimensioned parameter from the
interval $-mc^{2}<E<mc^{2}$ and $\mu_{n_{r}^{\prime}\kappa}^{(0)}(E)$
is a Sturmian eigenvalue; moreover, as we have previously assumed in
Section \ref{II}, it holds that $Z<\alpha^{-1}/2$. The reader should
observe that this is the \emph{inverse\/} of the Sturmian eigenvalue
$\mu_{n_{r}^{\prime}\kappa}^{(0)}(E)$ which multiplies the Coulomb
potential in the lower diagonal term of the differential operator in
Eq.\ (\ref{4.10a}). Proceeding as in Ref.\ \cite{Szmy97}, with some
labor one finds that eigensolutions to the system (\ref{4.10}) are
\begin{equation}
\mu_{n_{r}^{\prime}\kappa}^{(0)}(E)=\frac{\varepsilon}{\alpha Z}
(|n_{r}^{\prime}|+\gamma_{\kappa}+N_{n_{r}^{\prime}\kappa}^{\prime})
\label{4.11}
\end{equation}
and
\begin{subequations}
\begin{eqnarray}
S_{n_{r}^{\prime}\kappa}^{(0)}(E,r) 
&=& \sqrt{\frac{4\pi\epsilon_{0}}{e^{2}}
\frac{\alpha|n_{r}^{\prime}|!(|n_{r}^{\prime}|+2\gamma_{\kappa})}
{2\varepsilon N_{n_{r}^{\prime}\kappa}^{\prime}
(N_{n_{r}^{\prime}\kappa}^{\prime}-\kappa)
\Gamma(|n_{r}^{\prime}|+2\gamma_{\kappa})}}\,
(2kr)^{\gamma_{\kappa}}\mathrm{e}^{-kr}
\nonumber \\
&& \times\left[L_{|n_{r}^{\prime}|-1}^{(2\gamma_{\kappa})}(2kr)
-\frac{N_{n_{r}^{\prime}\kappa}^{\prime}-\kappa}
{|n_{r}^{\prime}|+2\gamma_{\kappa}}
L_{|n_{r}^{\prime}|}^{(2\gamma_{\kappa})}(2kr)\right],
\label{4.12a}
\end{eqnarray}
\begin{eqnarray}
T_{n_{r}^{\prime}\kappa}^{(0)}(E,r) 
&=& -\sqrt{\frac{4\pi\epsilon_{0}}{e^{2}}
\frac{\alpha\varepsilon|n_{r}^{\prime}|!
(|n_{r}^{\prime}|+2\gamma_{\kappa})}
{2N_{n_{r}^{\prime}\kappa}^{\prime}
(N_{n_{r}^{\prime}\kappa}^{\prime}-\kappa)
\Gamma(|n_{r}^{\prime}|+2\gamma_{\kappa})}}\,
(2kr)^{\gamma_{\kappa}}\mathrm{e}^{-kr}
\nonumber \\
&& \times\left[L_{|n_{r}^{\prime}|-1}^{(2\gamma_{\kappa})}(2kr)
+\frac{N_{n_{r}^{\prime}\kappa}^{\prime}-\kappa}
{|n_{r}^{\prime}|+2\gamma_{\kappa}}
L_{|n_{r}^{\prime}|}^{(2\gamma_{\kappa})}(2kr)\right],
\label{4.12b}
\end{eqnarray}
\label{4.12}%
\end{subequations}
where
\begin{equation}
\varepsilon=\sqrt{\frac{mc^{2}-E}{mc^{2}+E}},
\qquad
k=\frac{\sqrt{(mc^{2})^{2}-E^{2}}}{c\hbar}
\label{4.13}
\end{equation}
and
\begin{equation}
N_{n_{r}^{\prime}\kappa}^{\prime}
=\pm\sqrt{n_{r}^{\prime\,2}+2|n_{r}^{\prime}|\gamma_{\kappa}
+\kappa^{2}}.
\label{4.14}
\end{equation}
In contrast to the case of the energy-spectral problem discussed in
Section \ref{II}, the Sturmian radial quantum number $n_{r}^{\prime}$
used here runs through \emph{all\/} integers, i.e.,
$n_{r}^{\prime}\in\mathbb{Z}$. The following sign convention is
adopted in Eq.\ (\ref{4.14}): one chooses the positive sign for
$n_{r}^{\prime}>0$ and the negative sign for $n_{r}^{\prime}<0$; if
$n_{r}^{\prime}=0$, then the positive sign is to be chosen for
$\kappa\leqslant-\frac{1}{2}$ and the negative one for
$\kappa\geqslant\frac{1}{2}$, i.e., it holds that
$N_{0\kappa}^{\prime}=-\kappa$.

The functions given in Eqs.\ (\ref{4.12a}) and (\ref{4.12b}) possess
the following generalized orthogonality properties:
\begin{subequations}
\begin{equation}
\int_{0}^{\infty}\mathrm{d}r\:\frac{Ze^{2}}{(4\pi\epsilon_{0})r}
\left[\mu_{n_{r}^{\prime}\kappa}^{(0)}(E)
S_{n_{r}^{\prime}\kappa}^{(0)}(E,r)
S_{n_{r}^{\prime\prime}\kappa}^{(0)}(E,r)
-\mu_{n_{r}^{\prime\prime}\kappa}^{(0)\,-1}(E)
T_{n_{r}^{\prime}\kappa}^{(0)}(E,r)
T_{n_{r}^{\prime\prime}\kappa}^{(0)}(E,r)\right]
=\delta_{n_{r}^{\prime}n_{r}^{\prime\prime}} 
\label{4.15a}
\end{equation}
and
\begin{equation}
c\hbar k\int_{0}^{\infty}\mathrm{d}r
\left[\varepsilon S_{n_{r}^{\prime}\kappa}^{(0)}(E,r)
S_{n_{r}^{\prime\prime}\kappa}^{(0)}(E,r)
+\varepsilon^{-1}T_{n_{r}^{\prime}\kappa}^{(0)}(E,r)
T_{n_{r}^{\prime\prime}\kappa}^{(0)}(E,r)\right]
=\delta_{n_{r}^{\prime}n_{r}^{\prime\prime}}.
\label{4.15b}
\end{equation}
\label{4.15}%
\end{subequations}
Moreover, they obey the generalized closure relations
\begin{subequations}
\begin{equation}
\frac{Ze^{2}}{(4\pi\epsilon_{0})r}
\sum_{n_{r}^{\prime}=-\infty}^{\infty}
\left(
\begin{array}{c}
\mu_{n_{r}^{\prime}\kappa}^{(0)}(E)
S_{n_{r}^{\prime}\kappa}^{(0)}(E,r) \\*[1ex]
T_{n_{r}^{\prime}\kappa}^{(0)}(E,r)
\end{array}
\right)
\left(
\begin{array}{cc}
S_{n_{r}^{\prime}\kappa}^{(0)}(E,r') & 
-\mu_{n_{r}^{\prime}\kappa}^{(0)\,-1}(E)
T_{n_{r}^{\prime}\kappa}^{(0)}(E,r')
\end{array}
\right)
=\delta(r-r')I
\label{4.16a}
\end{equation}
and
\begin{equation}
c\hbar k\sum_{n_{r}^{\prime}=-\infty}^{\infty}
\left(
\begin{array}{c}
S_{n_{r}^{\prime}\kappa}^{(0)}(E,r) \\*[1ex]
T_{n_{r}^{\prime}\kappa}^{(0)}(E,r)
\end{array}
\right)
\left(
\begin{array}{cc}
\varepsilon S_{n_{r}^{\prime}\kappa}^{(0)}(E,r') & 
\varepsilon^{-1}T_{n_{r}^{\prime}\kappa}^{(0)}(E,r')
\end{array}
\right)
=\delta(r-r')I.
\label{4.16b}
\end{equation}
\label{4.16}%
\end{subequations}
It follows from Eqs.\ (\ref{4.11}) and (\ref{2.19}) that in the limit
$E\to E_{n\kappa}^{(0)}$ the Sturmian eigenvalue
$\mu_{n_{r}^{\prime}\kappa}^{(0)}(E)$, with nonnegative
$n_{r}^{\prime}\equiv n_{r}=n-|\kappa|-1/2$ [cf.\ Eq.\ (\ref{2.12})],
becomes equal to unity:
\begin{equation}
\mu_{n_{r}\kappa}^{(0)}(E_{n\kappa}^{(0)})=1
\qquad 
\left(n_{r}=n-|\kappa|-{\textstyle\frac{1}{2}}\geqslant
\left\{
\begin{array}{lll}
0 & \textrm{for} & \kappa\leqslant-\frac{1}{2} \\
1 & \textrm{for} & \kappa\geqslant\frac{1}{2}
\end{array}
\right.
\right).
\label{4.17}
\end{equation}
In the same limit and under the same restraint on $n_{r}^{\prime}$,
the Sturmian functions $S_{n_{r}^{\prime}\kappa}^{(0)}(E,r)$ and
$T_{n_{r}^{\prime}\kappa}^{(0)}(E,r)$ become
\begin{eqnarray}
&& S_{n_{r}\kappa}^{(0)}(E_{n\kappa}^{(0)},r)
=\frac{N_{n_{r}\kappa}}{Z}
\sqrt{\frac{(4\pi\epsilon_{0})a_{0}}{e^{2}}}\,P_{n\kappa}^{(0)}(r),
\qquad
T_{n_{r}\kappa}^{(0)}(E_{n\kappa}^{(0)},r)
=\frac{N_{n_{r}\kappa}}{Z}
\sqrt{\frac{(4\pi\epsilon_{0})a_{0}}{e^{2}}}\,Q_{n\kappa}^{(0)}(r)
\nonumber \\
&& \hspace*{15em} 
\left(n_{r}=n-|\kappa|-{\textstyle\frac{1}{2}}\geqslant
\left\{
\begin{array}{lll}
0 & \textrm{for} & \kappa\leqslant-\frac{1}{2} \\
1 & \textrm{for} & \kappa\geqslant\frac{1}{2}
\end{array}
\right.
\right).
\label{4.18}
\end{eqnarray}

The radial Dirac--Coulomb Green function $G_{\kappa}^{(0)}(E,r,r')$ is
defined as that particular solution to the inhomogeneous equation
\begin{subequations}
\begin{equation}
\big[H_{\kappa}^{(0)}(r)-E\big]G_{\kappa}^{(0)}(E,r,r')
=\delta(r-r')I
\qquad (-mc^{2}<E<mc^{2}),
\label{4.19a}
\end{equation}
which obeys the boundary conditions
\begin{equation}
G_{\kappa}^{(0)}(E,r,r')
\stackrel{r\to0}{\longrightarrow}0,
\qquad
G_{\kappa}^{(0)}(E,r,r')
\stackrel{r\to\infty}{\longrightarrow}0.
\label{4.19b}
\end{equation}
\label{4.19}%
\end{subequations}
One may seek $G_{\kappa}^{(0)}(E,r,r')$ in the form of the Sturmian
series
\begin{equation}
G_{\kappa}^{(0)}(E,r,r')
=\sum_{n_{r}^{\prime}=-\infty}^{\infty}
\left(
\begin{array}{c}
S_{n_{r}^{\prime}\kappa}^{(0)}(E,r) \\*[1ex]
T_{n_{r}^{\prime}\kappa}^{(0)}(E,r)
\end{array}
\right)
C_{n_{r}^{\prime}\kappa}^{(0)}(E,r').
\label{4.20}
\end{equation}
To determine the coefficients $C_{n_{r}^{\prime}\kappa}^{(0)}(E,r')$,
we insert the expansion (\ref{4.20}) into Eq.\ (\ref{4.19a}), then
exploit Eqs.\ (\ref{2.14}) and (\ref{4.10a}), premultiply the
resulting equation with
$\left(\begin{array}{cc}\mu_{n_{r}^{\prime\prime}\kappa}^{(0)}(E)
S_{n_{r}^{\prime\prime}\kappa}^{(0)}(E,r) &
T_{n_{r}^{\prime\prime}\kappa}^{(0)}(E,r)\end{array}\right)$, and
integrate with respect to $r$ over the interval $[0,\infty)$. With the
use of the orthogonality relation (\ref{4.15a}), this eventually
yields
\begin{equation}
C_{n_{r}^{\prime}\kappa}^{(0)}(E,r')
=\frac{1}{\mu_{n_{r}^{\prime}\kappa}^{(0)}(E)-1}
\left(
\begin{array}{cc}
\mu_{n_{r}^{\prime}\kappa}^{(0)}(E)
S_{n_{r}^{\prime}\kappa}^{(0)}(E,r') & 
T_{n_{r}^{\prime}\kappa}^{(0)}(E,r')
\end{array}
\right),
\label{4.21}
\end{equation}
and consequently the explicit form of the Sturmian expansion of the
radial Dirac--Coulomb Green function $G_{\kappa}^{(0)}(E,r,r')$ is
found to be
\begin{equation}
G_{\kappa}^{(0)}(E,r,r')
=\sum_{n_{r}^{\prime}=-\infty}^{\infty}
\frac{1}{\mu_{n_{r}^{\prime}\kappa}^{(0)}(E)-1}
\left(
\begin{array}{c}
S_{n_{r}^{\prime}\kappa}^{(0)}(E,r) \\*[1ex]
T_{n_{r}^{\prime}\kappa}^{(0)}(E,r)
\end{array}
\right)
\left(
\begin{array}{cc}
\mu_{n_{r}^{\prime}\kappa}^{(0)}(E)
S_{n_{r}^{\prime}\kappa}^{(0)}(E,r') & 
T_{n_{r}^{\prime}\kappa}^{(0)}(E,r')
\end{array}
\right).
\label{4.22}
\end{equation}

We are now ready to accomplish the task to determine the Sturmian
series representation of the generalized radial Dirac--Coulomb Green
function $\hat{G}_{n\kappa}^{(0)}(r,r')$. It is evident from
Eqs.\ (\ref{4.4}) and (\ref{4.19}) that the relationship between
$\hat{G}_{n\kappa}^{(0)}(r,r')$ and $G_{\kappa}^{(0)}(E,r,r')$
is
\begin{equation}
\hat{G}_{n\kappa}^{(0)}(r,r')=\lim_{E\to E_{n\kappa}^{(0)}}
\left[G_{\kappa}^{(0)}(E,r,r')
-\frac{\psi_{n\kappa}^{(0)}(r)\psi_{n\kappa}^{(0)\textrm{T}}(r')}
{E_{n\kappa}^{(0)}-E}\right].
\label{4.23}
\end{equation}
Upon exploiting the de l'Hospital rule, Eq.\ (\ref{4.23}) may be
rewritten as
\begin{equation}
\hat{G}_{n\kappa}^{(0)}(r,r')=\lim_{E\to E_{n\kappa}^{(0)}}
\left[\frac{\partial}{\partial E}\big(E-E_{n\kappa}^{(0)}\big)
G_{\kappa}^{(0)}(E,r,r')\right].
\label{4.24}
\end{equation}
If the expansion (\ref{4.22}) is plugged into the right-hand side of
Eq.\ (\ref{4.24}), with the aid of the identities
\begin{subequations}
\begin{equation}
\frac{\partial S_{n_{r}^{\prime}\kappa}^{(0)}(E,r)}{\partial E}
=-\frac{E}{(mc^{2})^{2}-E^{2}}
\left[r\frac{\mathrm{d}S_{n_{r}^{\prime}\kappa}^{(0)}(E,r)}
{\mathrm{d}r}
-\frac{mc^{2}}{2E}S_{n_{r}^{\prime}\kappa}^{(0)}(E,r)\right]
\label{4.25a}
\end{equation}
and
\begin{equation}
\frac{\partial T_{n_{r}^{\prime}\kappa}^{(0)}(E,r)}{\partial E}
=-\frac{E}{(mc^{2})^{2}-E^{2}}
\left[r\frac{\mathrm{d}T_{n_{r}^{\prime}\kappa}^{(0)}(E,r)}
{\mathrm{d}r}
+\frac{mc^{2}}{2E}T_{n_{r}^{\prime}\kappa}^{(0)}(E,r)\right],
\label{4.25b}
\end{equation}
\label{4.25}%
\end{subequations}
as well as of the relations
\begin{subequations}
\begin{equation}
\frac{E-E_{n\kappa}^{(0)}}{\mu_{n_{r}\kappa}^{(0)}(E)-1}
=\varepsilon_{n\kappa}^{(0)}
\frac{E-E_{n\kappa}^{(0)}}{\varepsilon-\varepsilon_{n\kappa}^{(0)}}
=-\frac{\varepsilon_{n\kappa}^{(0)}
(\varepsilon+\varepsilon_{n\kappa}^{(0)})
(mc^{2}+E)(mc^{2}+E_{n\kappa}^{(0)})}{2mc^{2}},
\label{4.26a}
\end{equation}
\begin{equation}
\lim_{E\to E_{n\kappa}^{(0)}}
\frac{E-E_{n\kappa}^{(0)}}{\mu_{n_{r}\kappa}^{(0)}(E)-1}
=-\frac{(mc^{2})^{2}-(E_{n\kappa}^{(0)})^{2}}{mc^{2}},
\label{4.26b}
\end{equation}
\begin{equation}
\lim_{E\to E_{n\kappa}^{(0)}}\frac{\partial}{\partial E}
\frac{E-E_{n\kappa}^{(0)}}{\mu_{n_{r}\kappa}^{(0)}(E)-1}
=\frac{2E_{n\kappa}^{(0)}-mc^{2}}{2mc^{2}}
\label{4.26c}
\end{equation}
and
\begin{equation}
\lim_{E\to E_{n\kappa}^{(0)}}
\frac{E-E_{n\kappa}^{(0)}}{\mu_{n_{r}\kappa}^{(0)}(E)-1}
\frac{\partial\mu_{n_{r}\kappa}^{(0)}(E)}{\partial E}=1,
\label{4.26d}
\end{equation}
\label{4.26}%
\end{subequations}
where $n_{r}$ is related to $n$ through Eq.\ (\ref{2.12}) and where
\begin{equation}
\varepsilon_{n\kappa}^{(0)}
=\sqrt{\frac{mc^{2}-E_{n\kappa}^{(0)}}{mc^{2}+E_{n\kappa}^{(0)}}}
=\frac{\alpha Z}{n_{r}+\gamma_{\kappa}+N_{n_{r}\kappa}},
\label{4.27}
\end{equation}
we find that the sought Sturmian expansion of
$\hat{G}_{n\kappa}^{(0)}(r,r')$ is
\begin{allowdisplaybreaks}
\begin{eqnarray}
\hat{G}_{n\kappa}^{(0)}(r,r')
&=& \sum_{\substack{n_{r}^{\prime}=-\infty \\ 
(n_{r}^{\prime}\neq n_{r})}}^{\infty}
\frac{1}{\mu_{n_{r}^{\prime}\kappa}^{(0)}(E_{n\kappa}^{(0)})-1}
\nonumber \\
&& \quad 
\times\left(
\begin{array}{c}
S_{n_{r}^{\prime}\kappa}^{(0)}(E_{n\kappa}^{(0)},r) \\*[1ex]
T_{n_{r}^{\prime}\kappa}^{(0)}(E_{n\kappa}^{(0)},r)
\end{array}
\right)
\left(
\begin{array}{cc}
\mu_{n_{r}^{\prime}\kappa}^{(0)}(E_{n\kappa}^{(0)})
S_{n_{r}^{\prime}\kappa}^{(0)}(E_{n\kappa}^{(0)},r') & 
T_{n_{r}^{\prime}\kappa}^{(0)}(E_{n\kappa}^{(0)},r')
\end{array}
\right)
\nonumber \\
&& +\,\frac{2E_{n\kappa}^{(0)}-mc^{2}}{2mc^{2}}
\left(
\begin{array}{c}
S_{n_{r}\kappa}^{(0)}(E_{n\kappa}^{(0)},r) \\*[1ex]
T_{n_{r}\kappa}^{(0)}(E_{n\kappa}^{(0)},r)
\end{array}
\right)
\left(
\begin{array}{cc}
S_{n_{r}\kappa}^{(0)}(E_{n\kappa}^{(0)},r') & 
T_{n_{r}\kappa}^{(0)}(E_{n\kappa}^{(0)},r')
\end{array}
\right)
\nonumber \\
&& +\left(
\begin{array}{c}
I_{n_{r}\kappa}^{(0)}(E_{n\kappa}^{(0)},r) \\*[1ex]
K_{n_{r}\kappa}^{(0)}(E_{n\kappa}^{(0)},r)
\end{array}
\right)
\left(
\begin{array}{cc}
S_{n_{r}\kappa}^{(0)}(E_{n\kappa}^{(0)},r') & 
T_{n_{r}\kappa}^{(0)}(E_{n\kappa}^{(0)},r')
\end{array}
\right)
\nonumber \\
&& +\left(
\begin{array}{c}
S_{n_{r}\kappa}^{(0)}(E_{n\kappa}^{(0)},r) \\*[1ex]
T_{n_{r}\kappa}^{(0)}(E_{n\kappa}^{(0)},r)
\end{array}
\right)
\left(
\begin{array}{cc}
J_{n_{r}\kappa}^{(0)}(E_{n\kappa}^{(0)},r') & 
K_{n_{r}\kappa}^{(0)}(E_{n\kappa}^{(0)},r')
\end{array}
\right),
\label{4.28}
\end{eqnarray}
\end{allowdisplaybreaks}
with
\begin{subequations}
\begin{eqnarray}
I_{n_{r}\kappa}^{(0)}(E_{n\kappa}^{(0)},r)
&=& \lim_{E\to E_{n\kappa}^{(0)}}
\left[\frac{E-E_{n\kappa}^{(0)}}
{\mu_{n_{r}\kappa}^{(0)}(E)-1}
\frac{\partial S_{n_{r}\kappa}^{(0)}(E,r)}{\partial E}\right]
\nonumber \\
&=& \frac{E_{n\kappa}^{(0)}}{mc^{2}}
\left[r\frac{\mathrm{d}S_{n_{r}\kappa}^{(0)}(E_{n\kappa}^{(0)},r)}
{\mathrm{d}r}-\frac{mc^{2}}{2E_{n\kappa}^{(0)}}
S_{n_{r}\kappa}^{(0)}(E_{n\kappa}^{(0)},r)\right],
\label{4.29a}
\end{eqnarray}
\begin{eqnarray}
J_{n_{r}\kappa}^{(0)}(E_{n\kappa}^{(0)},r)
&=& \lim_{E\to E_{n\kappa}^{(0)}}
\left[\frac{E-E_{n\kappa}^{(0)}}{\mu_{n_{r}\kappa}^{(0)}(E)-1}
\frac{\partial[\mu_{n_{r}\kappa}^{(0)}(E)S_{n_{r}\kappa}^{(0)}(E,r)]}
{\partial E}\right]
\nonumber \\
&=& I_{n_{r}\kappa}^{(0)}(E_{n\kappa}^{(0)},r)
+S_{n_{r}\kappa}^{(0)}(E_{n\kappa}^{(0)},r)
\nonumber \\
&=& \frac{E_{n\kappa}^{(0)}}{mc^{2}}
\left[r\frac{\mathrm{d}S_{n_{r}\kappa}^{(0)}(E_{n\kappa}^{(0)},r)}
{\mathrm{d}r}+\frac{mc^{2}}{2E_{n\kappa}^{(0)}}
S_{n_{r}\kappa}^{(0)}(E_{n\kappa}^{(0)},r)\right]
\label{4.29b}
\end{eqnarray}
and
\begin{eqnarray}
K_{n_{r}\kappa}^{(0)}(E_{n\kappa}^{(0)},r)
&=& \lim_{E\to E_{n\kappa}^{(0)}}
\left[\frac{E-E_{n\kappa}^{(0)}}{\mu_{n_{r}\kappa}^{(0)}(E)-1}
\frac{\partial T_{n_{r}\kappa}^{(0)}(E,r)}{\partial E}\right]
\nonumber \\
&=& \frac{E_{n\kappa}^{(0)}}{mc^{2}}
\left[r\frac{\mathrm{d}T_{n_{r}\kappa}^{(0)}(E_{n\kappa}^{(0)},r)}
{\mathrm{d}r}
+\frac{mc^{2}}{2E_{n\kappa}^{(0)}}
T_{n_{r}\kappa}^{(0)}(E_{n\kappa}^{(0)},r)\right].
\label{4.29c}
\end{eqnarray}
\label{4.29}%
\end{subequations}

With the expansion (\ref{4.28}) in hands, we may return to the problem
of evaluation of the second-order energy correction
$E_{n\kappa}^{(2)}$. Insertion of Eq.\ (\ref{4.28}) into Eq.\
(\ref{4.9}), followed by the use of Eqs.\ (\ref{4.29a})--(\ref{4.29c})
and of Eq.\ (\ref{4.18}), gives
\begin{eqnarray}
E_{n\kappa}^{(2)} &=& -\frac{1}{4}e^{2}c^{2}B^{2}
\sum_{\substack{n_{r}^{\prime}=-\infty \\ 
(n_{r}^{\prime}\neq n_{r})}}^{\infty}
\frac{1}{\mu_{n_{r}^{\prime}\kappa}^{(0)}(E_{n\kappa}^{(0)})-1}
\int_{0}^{\infty}\mathrm{d}r\:r\big[Q_{n\kappa}^{(0)}(r)
S_{n_{r}^{\prime}\kappa}^{(0)}(E_{n\kappa}^{(0)},r)
+P_{n\kappa}^{(0)}(r)
T_{n_{r}^{\prime}\kappa}^{(0)}(E_{n\kappa}^{(0)},r)\big]
\nonumber \\
&& \quad\times\int_{0}^{\infty}\mathrm{d}r'\:r'
\big[\mu_{n_{r}^{\prime}\kappa}^{(0)}(E_{n\kappa}^{(0)})
Q_{n\kappa}^{(0)}(r')
S_{n_{r}^{\prime}\kappa}^{(0)}(E_{n\kappa}^{(0)},r')
+P_{n\kappa}^{(0)}(r')
T_{n_{r}^{\prime}\kappa}^{(0)}(E_{n\kappa}^{(0)},r')\big]
\nonumber \\
&& +\,\frac{(4\pi\epsilon_{0})a_{0}B^{2}}{m}
\frac{N_{n_{r}\kappa}^{2}E_{n\kappa}^{(0)}}{Z^{2}}
\left[\int_{0}^{\infty}\mathrm{d}r\:r
P_{n\kappa}^{(0)}(r)Q_{n\kappa}^{(0)}(r)\right]^{2}.
\label{4.30}
\end{eqnarray}
The last integral on the right-hand side of Eq.\ (\ref{4.30}) is the
one displayed in Eq.\ (\ref{3.5}). The first and the second integrals
may be evaluated using Eqs.\ (\ref{2.24}) and (\ref{4.12}), with the
aid of the formula
\begin{eqnarray}
&& \hspace*{-5em}
\int_{0}^{\infty}\mathrm{d}x\:x^{\alpha+1}\mathrm{e}^{-x}
L_{n}^{(\alpha)}(x)L_{n'}^{(\alpha)}(x)
\nonumber \\
&=& -\frac{\Gamma(\alpha+n+2)}{n!}\delta_{n',n+1}
+\frac{(\alpha+2n+1)\Gamma(\alpha+n+1)}{n!}\delta_{n'n}
-\frac{\Gamma(\alpha+n+1)}{(n-1)!}\delta_{n',n-1}
\nonumber \\
&& \hspace*{25em} (\Real\alpha>-2),
\label{4.31}
\end{eqnarray}
which generalizes the one in Eq.\ (\ref{3.3}) and, similarly to the
latter, may be inferred from Eq.\ (\ref{3.4}). After much algebra, one
finds that
\begin{allowdisplaybreaks}
\begin{eqnarray}
&& \hspace*{-5em}
\int_{0}^{\infty}\mathrm{d}r\:r\big[Q_{n\kappa}^{(0)}(r)
S_{n_{r}^{\prime}\kappa}^{(0)}(E_{n\kappa}^{(0)},r)
+P_{n\kappa}^{(0)}(r)
T_{n_{r}^{\prime}\kappa}^{(0)}(E_{n\kappa}^{(0)},r)\big]
\nonumber \\
&=& -\,\frac{\alpha\sqrt{4\pi\epsilon_{0}}\,a_{0}^{3/2}}{4Ze}
N_{n_{r}\kappa}\sqrt{\frac{n_{r}!(n_{r}+2\gamma_{\kappa})
|n_{r}^{\prime}|!(|n_{r}^{\prime}|+2\gamma_{\kappa})}
{N_{n_{r}\kappa}(N_{n_{r}\kappa}-\kappa)
\Gamma(n_{r}+2\gamma_{\kappa})N_{n_{r}^{\prime}\kappa}^{\prime}
(N_{n_{r}^{\prime}\kappa}^{\prime}-\kappa)
\Gamma(|n_{r}^{\prime}|+2\gamma_{\kappa})}}
\nonumber \\
&& \times\bigg[\frac{(N_{n_{r}\kappa}-\kappa)
(N_{n_{r}^{\prime}\kappa}^{\prime}-N_{n_{r}\kappa}-2\kappa)
\Gamma(n_{r}+2\gamma_{\kappa})}{n_{r}!}
\delta_{|n_{r}^{\prime}|,n_{r}+1}
\nonumber \\
&& \quad +\,\frac{4(n_{r}+\gamma_{\kappa})
\Gamma(n_{r}+2\gamma_{\kappa})}{(n_{r}-1)!}
\delta_{n_{r}^{\prime},-n_{r}}
\nonumber \\
&& \quad +\,\frac{
(N_{n_{r}^{\prime}\kappa}^{\prime}-\kappa)
(N_{n_{r}\kappa}-N_{n_{r}^{\prime}\kappa}^{\prime}-2\kappa)
\Gamma(n_{r}+2\gamma_{\kappa}-1)}
{(n_{r}-1)!}\delta_{|n_{r}^{\prime}|,n_{r}-1}\bigg]
\qquad (n_{r}^{\prime}\neq n_{r})
%\nonumber \\
%&&
\label{4.32}
\end{eqnarray}
\end{allowdisplaybreaks}
and
\begin{allowdisplaybreaks}
\begin{eqnarray}
&& \hspace*{-5em}
\int_{0}^{\infty}\mathrm{d}r\:
r\big[\mu_{n_{r}^{\prime}\kappa}^{(0)}(E_{n\kappa}^{(0)})
Q_{n\kappa}^{(0)}(r)
S_{n_{r}^{\prime}\kappa}^{(0)}(E_{n\kappa}^{(0)},r)
+P_{n\kappa}^{(0)}(r)
T_{n_{r}^{\prime}\kappa}^{(0)}(E_{n\kappa}^{(0)},r)\big]
\nonumber \\
&=& -\,\frac{\alpha\sqrt{4\pi\epsilon_{0}}\,a_{0}^{3/2}}{8Ze}
N_{n_{r}\kappa}
\big[\mu_{n_{r}^{\prime}\kappa}^{(0)}(E_{n\kappa}^{(0)})-1\big]
\nonumber \\
&& \times\,\sqrt{\frac{n_{r}!(n_{r}+2\gamma_{\kappa})
|n_{r}^{\prime}|!(|n_{r}^{\prime}|+2\gamma_{\kappa})}
{N_{n_{r}\kappa}(N_{n_{r}\kappa}-\kappa)
\Gamma(n_{r}+2\gamma_{\kappa})N_{n_{r}^{\prime}\kappa}^{\prime}
(N_{n_{r}^{\prime}\kappa}^{\prime}-\kappa)
\Gamma(|n_{r}^{\prime}|+2\gamma_{\kappa})}}
\nonumber \\
&& \times\,\bigg\{
-\,\frac{(N_{n_{r}\kappa}-\kappa)\Gamma(n_{r}+2\gamma_{\kappa}+2)}
{n_{r}!(n_{r}+2\gamma_{\kappa})}\delta_{|n_{r}^{\prime}|,n_{r}+2}
\nonumber \\
&& \quad+\,\frac{2(N_{n_{r}\kappa}-\kappa)[2n_{r}+2\gamma_{\kappa}+1
-\kappa(N_{n_{r}\kappa}+N_{n_{r}^{\prime}\kappa}^{\prime})]
\Gamma(n_{r}+2\gamma_{\kappa})}{n_{r}!}
\delta_{|n_{r}^{\prime}|,n_{r}+1}
\nonumber \\
&& \quad-\,\frac{2[N_{n_{r}\kappa}^{2}+2(n_{r}+\gamma_{\kappa})^{2}]
\Gamma(n_{r}+2\gamma_{\kappa})}{N_{n_{r}\kappa}(n_{r}-1)!}
\delta_{n_{r}^{\prime},-n_{r}}
\nonumber \\
&& \quad-\,\frac{2(N_{n_{r}^{\prime}\kappa}^{\prime}-\kappa)
[2n_{r}+2\gamma_{\kappa}-1
-\kappa(N_{n_{r}\kappa}+N_{n_{r}^{\prime}\kappa}^{\prime})]
\Gamma(n_{r}+2\gamma_{\kappa}-1)}{(n_{r}-1)!}
\delta_{|n_{r}^{\prime}|,n_{r}-1}
\nonumber \\
&& \quad+\,\frac{(N_{n_{r}^{\prime}\kappa}^{\prime}-\kappa)
\Gamma(n_{r}+2\gamma_{\kappa})}{(n_{r}-2)!(n_{r}+2\gamma_{\kappa}-2)}
\delta_{|n_{r}^{\prime}|,n_{r}-2}\bigg\}
\qquad (n_{r}^{\prime}\neq n_{r}).
\label{4.33}
\end{eqnarray}
\end{allowdisplaybreaks}
On the path to Eq.\ (\ref{4.33}), we have made use of the identity
\begin{equation}
\frac{\mu_{n_{r}^{\prime}\kappa}^{(0)}(E_{n\kappa}^{(0)})+1}
{\mu_{n_{r}^{\prime}\kappa}^{(0)}(E_{n\kappa}^{(0)})-1}
=\left\{
\begin{array}{lcl}
\displaystyle
\frac{N_{n_{r}^{\prime}\kappa}^{\prime}+N_{n_{r}\kappa}}
{|n_{r}^{\prime}|-n_{r}}
&& \textrm{for $|n_{r}^{\prime}|\neq n_{r}$} \\*[2ex]
\displaystyle
-\frac{n_{r}+\gamma_{\kappa}}{N_{n_{r}\kappa}}
&& \textrm{for $n_{r}^{\prime}=-n_{r}$},
\end{array}
\right.
\label{4.34}
\end{equation}
which follows from the definitions (\ref{4.11}) and (\ref{2.19}).
Combining Eqs.\ (\ref{4.30}), (\ref{4.32}), (\ref{4.33}) and
(\ref{3.5}), after tedious algebraic manipulations, one eventually
arrives at the sought final result for the second-order energy
correction
\begin{eqnarray}
E_{n\kappa}^{(2)} &=& \frac{1}{16}
\bigg[-\kappa\left(3n_{r}^{2}+6n_{r}\gamma_{\kappa}
+4\gamma_{\kappa}^{2}-\kappa^2\right)
+\frac{n_{r}+\gamma_{\kappa}}{N_{n_{r}\kappa}}
\big(5n_{r}^{4}+20n_{r}^{3}\gamma_{\kappa}+n_{r}^{2}
+22n_{r}^{2}\gamma_{\kappa}^{2}+5n_{r}^{2}\kappa^{2}
\nonumber \\
&& +\,4n_{r}\gamma_{\kappa}^{3}+2n_{r}\gamma_{\kappa}
+10n_{r}\gamma_{\kappa}\kappa^{2}+4\gamma_{\kappa}^{2}\kappa^{2}
-2\kappa^{4}+\kappa^{2}\big)\bigg]Z^{-2}\frac{B^{2}}{B_{0}^{2}}
\frac{e^{2}}{(4\pi\epsilon_{0})a_{0}}.
\label{4.35}
\end{eqnarray}
The expression in Eq.\ (\ref{4.35}) simplifies considerably for the
states with $n_{r}=0$ (i.e., those with $\kappa=-n+\frac{1}{2}$), for
which it becomes
\begin{equation}
E_{n,-n+1/2}^{(2)}=\frac{1}{16}\big(n-{\textstyle\frac{1}{2}}\big)
(2\gamma_{n-1/2}+1)\Big[2\gamma_{n-1/2}^{2}+\gamma_{n-1/2}
-\left(n-{\textstyle\frac{1}{2}}\right)^{2}\Big]
Z^{-2}\frac{B^{2}}{B_{0}^{2}}\frac{e^{2}}{(4\pi\epsilon_{0})a_{0}}.
\label{4.36}
\end{equation}

In general, the relationship between the second-order energy
correction $E^{(2)}$ and the modulus of the induction vector
$\boldsymbol{B}$ characterizing the perturbing uniform magnetic field
may be written in the form
\begin{equation}
E^{(2)}=-\frac{1}{2}\left(\frac{\mu_{0}}{4\pi}\right)^{-1}\chi B^{2},
\label{4.37}
\end{equation}
where $\mu_{0}$ is the vacuum permeability. The factor of
proportionality, $\chi$, is the magnetizability (magnetic
susceptibility) of the system. Comparison of Eqs.\ (\ref{4.35}) and
(\ref{4.37}) shows that the magnetizability of the planar atom in the
state characterized by the quantum numbers $n$ and $\kappa$ is
\begin{eqnarray}
\chi_{n\kappa} &=& \frac{1}{8}
\bigg[\kappa\left(3n_{r}^{2}+6n_{r}\gamma_{\kappa}
+4\gamma_{\kappa}^{2}-\kappa^2\right)
-\frac{n_{r}+\gamma_{\kappa}}{N_{n_{r}\kappa}}
\big(5n_{r}^{4}+20n_{r}^{3}\gamma_{\kappa}+n_{r}^{2}
+22n_{r}^{2}\gamma_{\kappa}^{2}+5n_{r}^{2}\kappa^{2}
\nonumber \\
&& +\,4n_{r}\gamma_{\kappa}^{3}+2n_{r}\gamma_{\kappa}
+10n_{r}\gamma_{\kappa}\kappa^{2}+4\gamma_{\kappa}^{2}\kappa^{2}
-2\kappa^{4}+\kappa^{2}\big)\bigg]\frac{\alpha^{2}a_{0}^{3}}{Z^{2}}.
\label{4.38}
\end{eqnarray}
For states with $n_{r}=0$, Eq.\ (\ref{4.38}) yields
\begin{equation}
\chi_{n,-n+1/2}=-\frac{1}{8}\big(n-{\textstyle\frac{1}{2}}\big)
(2\gamma_{n-1/2}+1)\Big[2\gamma_{n-1/2}^{2}+\gamma_{n-1/2}
-\big(n-{\textstyle\frac{1}{2}}\big)^{2}\Big]
\frac{\alpha^{2}a_{0}^{3}}{Z^{2}}.
\label{4.39}
\end{equation}
In particular, for the ground state, for which $n=1$, one finds that
\begin{equation}
\chi_{1,-1/2}=-\frac{1}{64}(2\gamma_{1/2}+1)
\big(8\gamma_{1/2}^{2}+4\gamma_{1/2}-1\big)
\frac{\alpha^{2}a_{0}^{3}}{Z^{2}}.
\label{4.40}
\end{equation}

The formula in Eq.\ (\ref{4.38}) is a counterpart of the one derived
recently by Stefa{\'n}ska \cite{Stef15,Stef16a} for a
three-dimensional one-electron Dirac atom. It is interesting that the
result for the planar atom is expressible in terms of elementary
functions, while the one for an atom in three dimensions involves
irreducible generalized hypergeometric series ${}_{3}F_{2}$ with the
unit argument.
%
%%\newpage
%
\section{Recapitulation and discussion}
\label{V}
\setcounter{equation}{0}
The purpose of the present paper has been to analyze the influence of
a weak, static, uniform magnetic field on energy levels of a planar
Dirac one-electron atom. In the preceding sections, with the use of
the Rayleigh--Schr{\"o}dinger perturbation theory, we have found that
energy of the atomic state which evolves from the state $\Psi_{n\kappa
m_{\kappa}}^{(0)}(\boldsymbol{r})$ of the isolated atom is
\begin{equation}
E_{n\kappa m_{\kappa}}=E_{n\kappa}^{(0)}+E_{n\kappa m_{\kappa}}^{(1)}
+E_{n\kappa}^{(2)}+O(B^{3}/B_{0}^{3}),
\label{5.1}
\end{equation}
where
\begin{subequations}
\begin{equation}
E_{n\kappa}^{(0)}=mc^{2}+\varepsilon_{n\kappa}^{(0)}Z^{2}
\frac{e^{2}}{(4\pi\epsilon_{0})a_{0}},
\label{5.2a}
\end{equation}
\begin{equation}
E_{n\kappa m_{\kappa}}^{(1)}
=\varepsilon_{n\kappa m_{\kappa}}^{(1)}\frac{B}{B_{0}}
\frac{e^{2}}{(4\pi\epsilon_{0})a_{0}}
\label{5.2b}
\end{equation}
and
\begin{equation}
E_{n\kappa}^{(2)}=\varepsilon_{n\kappa}^{(2)}Z^{-2}
\frac{B^{2}}{B_{0}^{2}}\frac{e^{2}}{(4\pi\epsilon_{0})a_{0}}
\label{5.2c}
\end{equation}
\label{5.2}%
\end{subequations}
[here $B_{0}$ is the atomic unit of magnetic induction defined in Eq.\
(\ref{3.7})], with the dimensionless coefficients
$\varepsilon_{\ldots}^{(k)}$ given by
\begin{subequations}
\begin{equation}
\varepsilon_{n\kappa}^{(0)}=(\alpha Z)^{-2}
\left(\frac{n_{r}+\gamma_{\kappa}}{N_{n_{r}\kappa}}-1\right),
\label{5.3a}
\end{equation}
\begin{equation}
\varepsilon_{n\kappa m_{\kappa}}^{(1)}=-\frac{m_{\kappa}}{4\kappa}
\left[1-\frac{2\kappa(n_{r}+\gamma_{\kappa})}{N_{n_{r}\kappa}}\right]
\label{5.3b}
\end{equation}
and
\begin{eqnarray}
\varepsilon_{n\kappa}^{(2)} &=& \frac{1}{16}
\bigg[-\kappa\left(3n_{r}^{2}+6n_{r}\gamma_{\kappa}
+4\gamma_{\kappa}^{2}-\kappa^2\right)
+\frac{n_{r}+\gamma_{\kappa}}{N_{n_{r}\kappa}}
\big(5n_{r}^{4}+20n_{r}^{3}\gamma_{\kappa}+n_{r}^{2}
+22n_{r}^{2}\gamma_{\kappa}^{2}+5n_{r}^{2}\kappa^{2}
\nonumber \\
&& +\,4n_{r}\gamma_{\kappa}^{3}+2n_{r}\gamma_{\kappa}
+10n_{r}\gamma_{\kappa}\kappa^{2}+4\gamma_{\kappa}^{2}\kappa^{2}
-2\kappa^{4}+\kappa^{2}\big)\bigg].
\label{5.3c}
\end{eqnarray}
\label{5.3}%
\end{subequations}
In Tables \ref{T.2} and \ref{T.3}, we display explicit expressions for
the coefficients $\varepsilon_{n\kappa}^{(0)}$, $\varepsilon_{n\kappa
m_{\kappa}}^{(1)}$ and $\varepsilon_{n\kappa}^{(2)}$ for atomic states
with the principal quantum numbers $1\leqslant n\leqslant3$.
\begin{center}
[Place for Tables \ref{T.2} and \ref{T.3}]
\end{center}

The reader may wish to observe that with the use of the coefficient
$\varepsilon_{n\kappa}^{(2)}$, the magnetizabilities (\ref{4.38}) may
be written as
\begin{equation}
\chi_{n\kappa}=-2\varepsilon_{n\kappa}^{(2)}
\frac{\alpha^{2}a_{0}^{3}}{Z^{2}}.
\label{5.4}
\end{equation}

For test purposes, we used the expressions in Eqs.\
(\ref{5.1})--(\ref{5.3}) to compute numerical values of the
second-order perturbation-theory estimates of the eigenenergies
$E_{n\kappa m_{\kappa}}$ for the magnetic-field perturbed planar atom
in states with the principal quantum numbers $n=1$ and $n=2$. In the
weak-field limit, results have been found to be in an excellent
agreement with corresponding numerically exact values obtained by
A.\/~Poszwa (private communication), who used the method presented in
Ref.\ \cite{Posz11}.

Expanding the expressions in Eqs.\ (\ref{5.3a})--(\ref{5.3b}) in the
Maclaurin series in $\alpha Z$, and retaining terms of orders not
higher than quadratic in that variable, one finds the following
quasi-relativistic approximations to the coefficients
$\varepsilon_{\ldots}^{(k)}$:
\begin{subequations}
\begin{equation}
\varepsilon_{n\kappa}^{(0)}
=-\frac{1}{2\left(n-\frac{1}{2}\right)^{2}}
\left[1+(\alpha Z)^{2}\frac{1}{\left(n-\frac{1}{2}\right)^{2}}
\left(\frac{n-\frac{1}{2}}{|\kappa|}-\frac{3}{4}\right)\right]
+\mathrm{O}\big((\alpha Z)^{4}\big),
\label{5.5a}
\end{equation}
\begin{equation}
\varepsilon_{n\kappa m_{\kappa}}^{(1)}
=\left\{
\begin{array}{lcl}
{\displaystyle
\frac{m_{\kappa}(2\kappa-1)}{4\kappa}
\left[1-(\alpha Z)^{2}
\frac{\kappa}{(2\kappa-1)\left(n-\frac{1}{2}\right)^{2}}\right]}
+\mathrm{O}\big((\alpha Z)^{4}\big)
&& \textrm{for $\kappa\neq\frac{1}{2}$} \\*[5ex]
{\displaystyle
-(\alpha Z)^{2}\frac{m_{\kappa}}{4\left(n-\frac{1}{2}\right)^{2}}}
+\mathrm{O}\big((\alpha Z)^{4}\big)
&& \textrm{for $\kappa=\frac{1}{2}$}
\end{array}
\right.
\label{5.5b}
\end{equation}
and
\begin{eqnarray}
\varepsilon_{n\kappa}^{(2)} &=& \frac{1}{64}
\big(n-{\textstyle\frac{1}{2}}\big)^{2}
(20n^{2}-20n-12\kappa^{2}-12\kappa+9)
\nonumber \\
&& \times\,\left[1+(\alpha Z)^{2}
\frac{\beta_{n\kappa}^{(2)}}
{2|\kappa|\left(n-\frac{1}{2}\right)^{2}
(20n^{2}-20n-12\kappa^{2}-12\kappa+9)}\right]
+\mathrm{O}\big((\alpha Z)^{4}\big),
\nonumber \\
&&
\label{5.5c}
\end{eqnarray}
\label{5.5}%
\end{subequations}
where
\begin{eqnarray}
\beta_{n\kappa}^{(2)}
&=& -\,80n^{3}+120n^{2}+44n^{2}|\kappa|-68n+24n\kappa^{2}
+24n\kappa-44n|\kappa|-28\kappa^{2}|\kappa|+8\kappa|\kappa|
\nonumber \\
&& -\,12\kappa^{2}-12\kappa+15|\kappa|+14.
\label{5.6}
\end{eqnarray}
For the ground state (i.e., the one with $n=1$, $\kappa=-\frac{1}{2}$
and $m_{\kappa}=\pm\frac{1}{2}$), Eqs.\ (\ref{5.5a})--(\ref{5.5c})
become
\begin{subequations}
\begin{equation}
\varepsilon_{1,-1/2}^{(0)}=-2\left[1+(\alpha Z)^{2}\right]
+\mathrm{O}\big((\alpha Z)^{4}\big),
\label{5.7a}
\end{equation}
\begin{equation}
\varepsilon_{1,-1/2,m_{\kappa}}^{(1)}
=m_{\kappa}\left[1-(\alpha Z)^{2}\right]
+\mathrm{O}\big((\alpha Z)^{4}\big)
\label{5.7b}
\end{equation}
and
\begin{equation}
\varepsilon_{1,-1/2}^{(2)}
=\frac{3}{64}\left[1-5(\alpha Z)^{2}\right]
+\mathrm{O}\big((\alpha Z)^{4}\big).
\label{5.7c}
\end{equation}
\label{5.7}%
\end{subequations}

In the purely nonrelativistic limit, i.e., for $\alpha\to0$, Eqs.\
(\ref{5.5a})--(\ref{5.5c}) yield
\begin{subequations}
\begin{equation}
\varepsilon_{n\kappa}^{(0)}
\stackrel{c\to\infty}{\longrightarrow}
-\frac{1}{2\left(n-\frac{1}{2}\right)^{2}},
\label{5.8a}
\end{equation}
\begin{equation}
\varepsilon_{n\kappa m_{\kappa}}^{(1)}
\stackrel{c\to\infty}{\longrightarrow}
\left\{
\begin{array}{lcl}
{\displaystyle\frac{m_{\kappa}(2\kappa-1)}{4\kappa}}
&& \textrm{for $\kappa\neq\frac{1}{2}$} \\*[3ex]
0 && \textrm{for $\kappa=\frac{1}{2}$}
\end{array}
\right.
\label{5.8b}
\end{equation}
and
\begin{equation}
\varepsilon_{n\kappa}^{(2)}
\stackrel{c\to\infty}{\longrightarrow}
\frac{1}{64}\big(n-{\textstyle\frac{1}{2}}\big)^{2}
(20n^{2}-20n-12\kappa^{2}-12\kappa+9).
\label{5.8c}
\end{equation}
\label{5.8}%
\end{subequations}
To facilitate comparison of the above limits with results of direct
nonrelativistic calculations reported in Ref.\ \cite{Szmy18} (cf.\
also Ref.\ \cite{Fern18}), Eqs.\ (\ref{5.8b}) and (\ref{5.8c}) should
be transformed. To this end, in the case of Eq.\ (\ref{5.8b}) we
introduce two quantum numbers $m_{l}$ and $m_{s}$, relating them to
$\kappa$ and $m_{\kappa}$ in the following way:
\begin{equation}
m_{l}=m_{\kappa}+\frac{m_{\kappa}}{2\kappa},
\qquad
m_{s}=-\frac{m_{\kappa}}{2\kappa}
\label{5.9}
\end{equation}
\begin{center}
[Place for Table \ref{T.4}]
\end{center}
(cf.\ also Table \ref{T.4}). It is evident that $m_{s}=\pm1/2$ and
$m_{l}=m_{\kappa}\mp1/2$, and that relations inverse to those in Eq.\
(\ref{5.9}) are
\begin{equation}
\kappa=-\frac{1}{2}\left(1+\frac{m_{l}}{m_{s}}\right),
\qquad
m_{\kappa}=m_{l}+m_{s}.
\label{5.10}
\end{equation}
Insertion of the latter into Eq.\ (\ref{5.8b}) gives
\begin{equation}
\varepsilon_{n\kappa m_{\kappa}}^{(1)}
\stackrel{c\to\infty}{\longrightarrow}
\frac{1}{2}(m_{l}+2m_{s}).
\label{5.11}
\end{equation}
To transform Eq.\ (\ref{5.8c}), we observe that it holds that
\begin{equation}
\kappa(\kappa+1)=l^{2}-\frac{1}{4},
\label{5.12}
\end{equation}
where the nonnegative integer $l$ has been defined in Eq.\
(\ref{2.26}); the reader may also wish to verify that $l=|m_{l}|$.
Plugging Eq.\ (\ref{5.12}) into Eq.\ (\ref{5.8c}) casts the latter
into the form
\begin{equation}
\varepsilon_{n\kappa}^{(2)}
\stackrel{c\to\infty}{\longrightarrow}
\frac{1}{16}\big(n-{\textstyle\frac{1}{2}}\big)^{2}
(5n^{2}-5n-3l^{2}+3).
\label{5.13}
\end{equation}
The expressions on the right-hand sides of Eqs.\ (\ref{5.8a}) and
(\ref{5.13}) are exactly the same as those in Eqs.\ (73) and (75)
from Ref.\ \cite{Szmy18}, respectively, while the expression on the
right-hand side of Eq.\ (\ref{5.11}) is identical to the one which may
be inferred from Eqs.\ (71) and (83) in Ref.\ \cite{Szmy18}.

The quasi-relativistic approximations for the magnetizabilities
$\chi_{n\kappa}$ may be easily deduced from Eqs.\ (\ref{5.4}) and
(\ref{5.5c}). In the most interesting case of the ground state, one
finds that
\begin{equation}
\chi_{1,-1/2}=\left\{-\frac{3}{32}\left[1-5(\alpha Z)^{2}\right]
+\mathrm{O}\big((\alpha Z)^{4}\big)\right\}
\frac{\alpha^{2}a_{0}^{3}}{Z^{2}}.
\label{5.14}
\end{equation}
%
%\newpage
%
\section*{Acknowledgments}
I am grateful to Dr.\ Andrzej Poszwa for kindly supplying me with his
unpublished numerical results for energy levels of the planar Dirac
one-electron atom in the perpendicular magnetic field. I also thank
Dr.\ Patrycja Stefa{\'n}ska for commenting on the manuscript.
%
%\newpage
%
\appendix
\section{Appendix: The axial spinors}
\label{A}
\setcounter{equation}{0}
The axial (or cylindrical) spinors, introduced by Poszwa and Rutkowski
\cite{Posz10}, are two-component functions of the angular variable
$\varphi\in[0,2\pi)$ defined as
\begin{equation}
\Phi_{\kappa m_{\kappa}}(\varphi)
=\frac{1}{\sqrt{2\pi}}
\left(
\begin{array}{c}
\delta_{-\kappa,m_{\kappa}}\,
\mathrm{e}^{\mathrm{i}(m_{\kappa}-1/2)\varphi}
\\*[1ex]
\delta_{\kappa,m_{\kappa}}\,\,
\mathrm{e}^{\mathrm{i}(m_{\kappa}+1/2)\varphi}
\end{array}
\right)
\qquad
(\textrm{${\textstyle\kappa=\pm\frac{1}{2},\pm\frac{3}{2},
\pm\frac{5}{2},\ldots}$; $m_{\kappa}=\pm\kappa$}),
\label{A.1}
\end{equation}
or equivalently as
\begin{equation}
\Phi_{\kappa m_{\kappa}}(\varphi)
=\frac{1}{\sqrt{2\pi}}
\left(
\begin{array}{c}
\delta_{-\kappa,m_{\kappa}}\,
\mathrm{e}^{-\mathrm{i}(\kappa+1/2)\varphi}
\\*[1ex]
\delta_{\kappa,m_{\kappa}}\,
\mathrm{e}^{\mathrm{i}(\kappa+1/2)\varphi}
\end{array}
\right)
\qquad
(\textrm{${\textstyle\kappa=\pm\frac{1}{2},\pm\frac{3}{2},
\pm\frac{5}{2},\ldots}$; $m_{\kappa}=\pm\kappa$})
\label{A.2}
\end{equation}
(the quantum number $\kappa$ appearing in Eqs.\ (\ref{A.1}) and
(\ref{A.2}), and in the rest of the present paper, is defined with the
sign \emph{opposite\/} in relation to the one used in Refs.\
\cite{Posz10,Posz11,Posz14}). Explicit forms of the spinors
$\Phi_{\kappa m_{\kappa}}(\varphi)$ are thus
\begin{equation}
\Phi_{\kappa,-\kappa}(\varphi)
=\frac{1}{\sqrt{2\pi}}
\left(
\begin{array}{c}
\mathrm{e}^{-\mathrm{i}(\kappa+1/2)\varphi} \\
0
\end{array}
\right),
\qquad
\Phi_{\kappa\kappa}(\varphi)
=\frac{1}{\sqrt{2\pi}}
\left(
\begin{array}{c}
0 \\
\mathrm{e}^{\mathrm{i}(\kappa+1/2)\varphi}
\end{array}
\right).
\label{A.3}
\end{equation}
These functions are orthonormal in the sense of
\begin{equation}
\int_{0}^{2\pi}\mathrm{d}\varphi\:
\Phi_{\kappa m_{\kappa}}^{\dag}(\varphi)
\Phi_{\kappa'm_{\kappa}^{\prime}}(\varphi)
=\delta_{\kappa\kappa'}\delta_{m_{\kappa}m_{\kappa}^{\prime}}
\label{A.4}
\end{equation}
and form a set which is complete in the space of square-integrable
two-component spinor functions of $\varphi\in[0,2\pi)$; the
corresponding closure relation is
\begin{equation}
\sum_{\kappa=-\infty-1/2}^{+\infty+1/2}
\sum_{m_{\kappa}=\pm\kappa}\Phi_{\kappa m_{\kappa}}(\varphi)
\Phi_{\kappa m_{\kappa}}^{\dag}(\varphi')
=\delta(\varphi-\varphi')I,
\label{A.5}
\end{equation}
where $I$ is the $2\times2$ unit matrix.

The products of $\Phi_{\kappa m_{\kappa}}(\varphi)$ with
$\cos\varphi$ or $\sin\varphi$ have the expansions
\begin{subequations}
\begin{equation}
\cos\varphi\,\Phi_{\kappa m_{\kappa}}(\varphi)
=\frac{1}{2}\Phi_{\kappa+m_{\kappa}/\kappa,m_{\kappa}+1}(\varphi)
+\frac{1}{2}\Phi_{\kappa-m_{\kappa}/\kappa,m_{\kappa}-1}(\varphi)
\label{A.6a}
\end{equation}
and
\begin{equation}
\sin\varphi\,\Phi_{\kappa m_{\kappa}}(\varphi)
=\frac{1}{2\mathrm{i}}
\Phi_{\kappa+m_{\kappa}/\kappa,m_{\kappa}+1}(\varphi)
-\frac{1}{2\mathrm{i}}
\Phi_{\kappa-m_{\kappa}/\kappa,m_{\kappa}-1}(\varphi).
\label{A.6b}
\end{equation}
\label{A.6}%
\end{subequations}
This leads to the following integral formulas:
\begin{subequations}
\begin{eqnarray}
\int_{0}^{2\pi}\mathrm{d}\varphi\:\cos\varphi\,
\Phi_{\kappa m_{\kappa}}^{\dag}(\varphi)
\Phi_{\kappa'm_{\kappa}^{\prime}}(\varphi)
&=& \frac{1}{2}\delta_{m_{\kappa}/\kappa,m_{\kappa}^{\prime}/\kappa'}
(\delta_{m_{\kappa},m_{\kappa}^{\prime}+1}
+\delta_{m_{\kappa},m_{\kappa}^{\prime}-1})
\nonumber \\
&=& \frac{1}{2}\delta_{m_{\kappa}/\kappa,m_{\kappa}^{\prime}/\kappa'}
(\delta_{\kappa,\kappa'+1}+\delta_{\kappa,\kappa'-1})
\label{A.7a}
\end{eqnarray}
and
\begin{eqnarray}
\int_{0}^{2\pi}\mathrm{d}\varphi\:\sin\varphi\,
\Phi_{\kappa m_{\kappa}}^{\dag}(\varphi)
\Phi_{\kappa'm_{\kappa}^{\prime}}(\varphi)
&=& \frac{1}{2\mathrm{i}}\,
\delta_{m_{\kappa}/\kappa,m_{\kappa}^{\prime}/\kappa'}
(\delta_{m_{\kappa},m_{\kappa}^{\prime}+1}
-\delta_{m_{\kappa},m_{\kappa}^{\prime}-1})
\nonumber \\
&=& \frac{1}{2\mathrm{i}}\sgn(m_{\kappa}/\kappa)
\delta_{m_{\kappa}/\kappa,m_{\kappa}^{\prime}/\kappa'}
(\delta_{\kappa,\kappa'+1}-\delta_{\kappa,\kappa'-1}).
\nonumber \\
&&
\label{A.7b}
\end{eqnarray}
\label{A.7}
\end{subequations}

If $\sigma_{1}$, $\sigma_{2}$, $\sigma_{3}$ are the Pauli matrices
\begin{equation}
\sigma_{1}
=\left(
\begin{array}{cc}
0 & 1 \\
1 & 0
\end{array}
\right),
\qquad
\sigma_{2}
=\left(
\begin{array}{cc}
0 & -\mathrm{i} \\
\mathrm{i} & 0
\end{array}
\right),
\qquad
\sigma_{3}
=\left(
\begin{array}{cc}
1 & 0 \\
0 & -1
\end{array}
\right),
\label{A.8}
\end{equation}
then it holds that
\begin{subequations}
\begin{equation}
\sigma_{1}\Phi_{\kappa m_{\kappa}}(\varphi)
=\Phi_{-\kappa-1,m_{\kappa}+m_{\kappa}/\kappa}(\varphi),
\label{A.9a}
\end{equation}
\begin{equation}
\sigma_{2}\Phi_{\kappa m_{\kappa}}(\varphi)
=-\frac{\mathrm{i}m_{\kappa}}{\kappa}
\Phi_{-\kappa-1,m_{\kappa}+m_{\kappa}/\kappa}(\varphi)
\label{A.9b}
\end{equation}
and
\begin{equation}
\sigma_{3}\Phi_{\kappa m_{\kappa}}(\varphi)
=-\frac{m_{\kappa}}{\kappa}\Phi_{\kappa m_{\kappa}}(\varphi).
\label{A.9c}
\end{equation}
\label{A.9}%
\end{subequations}

Let $\boldsymbol{n}_{x}$ and $\boldsymbol{n}_{y}$ be the unit vectors
of a planar Cartesian coordinate system $\{x,y\}$ and let
\begin{equation} 
\boldsymbol{n}_{r}=\boldsymbol{n}_{x}\cos\varphi
+\boldsymbol{n}_{y}\sin\varphi, \qquad
\boldsymbol{n}_{\varphi}=-\boldsymbol{n}_{x}\sin\varphi
+\boldsymbol{n}_{y}\cos\varphi
\label{A.10}
\end{equation}
be the unit vectors of a polar coordinate system $\{r,\varphi\}$ with
the same origin. It holds that
\begin{subequations}
\begin{equation}
\boldsymbol{n}_{r}\cdot\boldsymbol{\sigma}
\Phi_{\kappa m_{\kappa}}(\varphi)
=\Phi_{-\kappa m_{\kappa}}(\varphi)
\label{A.11a}
\end{equation}
and
\begin{equation}
\boldsymbol{n}_{\varphi}\cdot\boldsymbol{\sigma}
\Phi_{\kappa m_{\kappa}}(\varphi)
=-\frac{\mathrm{i}m_{\kappa}}{\kappa}
\Phi_{-\kappa m_{\kappa}}(\varphi),
\label{A.11b}
\end{equation}
\label{A.11}%
\end{subequations}
where
\begin{equation}
\boldsymbol{\sigma}
=\sigma_{1}\boldsymbol{n}_{x}+\sigma_{2}\boldsymbol{n}_{y}.
\label{A.12}
\end{equation}
The reader may wish to observe that results for the expressions
$(\boldsymbol{n}_{z}\times\boldsymbol{n}_{\varphi})
\cdot\boldsymbol{\sigma}\Phi_{\kappa m_{\kappa}}(\varphi)$ and
\linebreak $(\boldsymbol{n}_{z}\times\boldsymbol{n}_{r})
\cdot\boldsymbol{\sigma}\Phi_{\kappa m_{\kappa}}(\varphi)$, where
\begin{equation}
\boldsymbol{n}_{z}=\boldsymbol{n}_{x}\times\boldsymbol{n}_{y},
\label{A.13}
\end{equation}
may be deduced immediately from Eqs.\ (\ref{A.11a}) and (\ref{A.11b}),
respectively, since one has
\begin{equation}
\boldsymbol{n}_{z}\times\boldsymbol{n}_{\varphi}=-\boldsymbol{n}_{r},
\qquad
\boldsymbol{n}_{z}\times\boldsymbol{n}_{r}=\boldsymbol{n}_{\varphi}.
\label{A.14}
\end{equation}

Equation (\ref{A.9c}) expresses the fact that the axial spinors are
eigenvectors of the Pauli matrix $\sigma_{3}$. They also appear to be
simultaneous eigenfunctions of the three operators
\begin{equation}
\Lambda=-\mathrm{i}\frac{\partial}{\partial\varphi},
\qquad
J=\Lambda+\frac{1}{2}\sigma_{3},
\qquad
K=-\left(\sigma_{3}\Lambda+\frac{1}{2}I\right)
=-\sigma_{3}J,
\label{A.15}
\end{equation}
as it holds that
\begin{subequations}
\begin{equation}
\Lambda\Phi_{\kappa m_{\kappa}}(\varphi)
=\frac{m_{\kappa}}{\kappa}\left(\kappa+\frac{1}{2}\right)
\Phi_{\kappa m_{\kappa}}(\varphi),
\label{A.16a}
\end{equation}
\begin{equation}
J\Phi_{\kappa m_{\kappa}}(\varphi)
=m_{\kappa}\Phi_{\kappa m_{\kappa}}(\varphi)
\label{A.16b}
\end{equation}
and
\begin{equation}
K\Phi_{\kappa m_{\kappa}}(\varphi)
=\kappa\Phi_{\kappa m_{\kappa}}(\varphi).
\label{A.16c}
\end{equation}
\label{A.16}%
\end{subequations}

The result of the action of the operator
\begin{equation}
\boldsymbol{\sigma}\cdot\boldsymbol{\nabla}
=\boldsymbol{n}_{r}\cdot\boldsymbol{\sigma}
\frac{\partial}{\partial r}+\frac{1}{r}
\boldsymbol{n}_{\varphi}\cdot\boldsymbol{\sigma}
\frac{\partial}{\partial\varphi}
\label{A.17}
\end{equation}
on the product $F(r)\Phi_{\kappa m_{\kappa}}(\varphi)$ is
\begin{equation}
\boldsymbol{\sigma}\cdot\boldsymbol{\nabla}
F(r)\Phi_{\kappa m_{\kappa}}(\varphi)
=\left(\frac{\partial}{\partial r}
+\frac{\kappa+\frac{1}{2}}{r}\right)
F(r)\Phi_{-\kappa m_{\kappa}}(\varphi).
\label{A.18}
\end{equation}
%
%\newpage
%

%
\newpage
\begin{table}[t] 
\caption{Relativistic quantum numbers and the spectroscopic
designation for selected states of the planar Dirac one-electron atom
(after Ref.\ \cite{Posz10}, except for the quantum number $\kappa$
which in the present paper is defined with the sign \emph{opposite\/}
in relation to the one used in Refs.\ \cite{Posz10,Posz11,Posz14}).}
\label{T.1}
\begin{center}
\begin{tabular}{ccccccccc}
\hline \\*[-0.5ex]
\multicolumn{8}{}{} & Spectroscopic \\
$n$ && $n_{r}$ && $\kappa$ && $l=|\kappa+\frac{1}{2}|$ && notation \\
\multicolumn{8}{}{} & $nl_{|\kappa|}$ \\*[1ex]
\hline \\*[-0.5ex]
1 && 0 && $-\frac{1}{2}$           && 0 && 1s$_{1/2}$ \\*[1ex]
2 && 1 && $-\frac{1}{2}$           && 0 && 2s$_{1/2}$ \\*[1ex]
2 && 1 && $\phantom{-}\frac{1}{2}$ && 1 && 2p$_{1/2}$ \\*[1ex]
2 && 0 && $-\frac{3}{2}$           && 1 && 2p$_{3/2}$ \\*[1ex]
3 && 2 && $-\frac{1}{2}$           && 0 && 3s$_{1/2}$ \\*[1ex]
3 && 2 && $\phantom{-}\frac{1}{2}$ && 1 && 3p$_{1/2}$ \\*[1ex]
3 && 1 && $-\frac{3}{2}$           && 1 && 3p$_{3/2}$ \\*[1ex]
3 && 1 && $\phantom{-}\frac{3}{2}$ && 2 && 3d$_{3/2}$ \\*[1ex]
3 && 0 && $-\frac{5}{2}$           && 2 && 3d$_{5/2}$ \\*[1ex]
\hline
\end{tabular}
\end{center}
\end{table}
%
%\newpage
%
\begin{landscape} 
\begin{table}[h] 

\caption{Explicit forms of the coefficients
$\varepsilon_{n\kappa}^{(0)}$ [defined in Eq.\ (\ref{5.3a})] and
$\varepsilon_{n\kappa m_{\kappa}}^{(1)}$ [defined in Eq.\
(\ref{5.3b})] for atomic states with the principal quantum numbers
$1\leqslant n\leqslant3$. The symbol $\gamma_{\kappa}$ has been
defined in Eq.\ (\ref{2.21}).}

\label{T.2}
\begin{center}
\begin{tabular}{lclccclcc}
\hline \\*[-0.5ex]
\multicolumn{1}{c}{Atomic} && 
\multicolumn{3}{c}{$\varepsilon_{n\kappa}^{(0)}$} &&
\multicolumn{3}{c}{$\varepsilon_{n\kappa m_{\kappa}}^{(1)}$} 
\\*[0.5ex]
\cline{3-5} \cline{7-9}\\*[-1.25ex]
\multicolumn{1}{c}{state} && 
\multicolumn{1}{c}{Exact} && Nonrelativistic limit &&
\multicolumn{1}{c}{Exact} && Nonrelativistic limit \\*[1ex]
\hline \\*[-0.5ex]
1s$_{1/2}$ && 
$(\alpha Z)^{-2}(2\gamma_{1/2}-1)$ && 
$-2$ && 
$\displaystyle\frac{1}{4}\sgn(m_{\kappa})(2\gamma_{1/2}+1)$ && 
$\displaystyle\frac{1}{2}\sgn(m_{\kappa})$ \\*[4ex]
2s$_{1/2}$ && 
$\displaystyle(\alpha Z)^{-2}\left[\frac{2(\gamma_{1/2}+1)}
{\sqrt{8\gamma_{1/2}+5}}-1\right]$ &&
$\displaystyle-\frac{2}{9}$ && 
$\displaystyle\frac{1}{4}\sgn(m_{\kappa})
\left[\frac{2(\gamma_{1/2}+1)}{\sqrt{8\gamma_{1/2}+5}}+1\right]$ && 
$\displaystyle\frac{1}{2}\sgn(m_{\kappa})$ \\*[4ex]
2p$_{1/2}$ && 
$\displaystyle(\alpha Z)^{-2}\left[\frac{2(\gamma_{1/2}+1)}
{\sqrt{8\gamma_{1/2}+5}}-1\right]$ && 
$\displaystyle-\frac{2}{9}$ &&
$\displaystyle\frac{1}{4}\sgn(m_{\kappa})
\left[\frac{2(\gamma_{1/2}+1)}{\sqrt{8\gamma_{1/2}+5}}-1\right]$ && 
0 \\*[4ex]
2p$_{3/2}$ && 
$\displaystyle(\alpha Z)^{-2}\left(\frac{2}{3}\gamma_{3/2}-1\right)$ && 
$\displaystyle-\frac{2}{9}$ &&
$\displaystyle\frac{1}{4}\sgn(m_{\kappa})(2\gamma_{3/2}+1)$ && 
$\sgn(m_{\kappa})$ \\*[4ex]
3s$_{1/2}$ && 
$\displaystyle(\alpha Z)^{-2}\left[\frac{2(\gamma_{1/2}+2)}
{\sqrt{16\gamma_{1/2}+17}}-1\right]$ && 
$\displaystyle-\frac{2}{25}$ &&
$\displaystyle\frac{1}{4}\sgn(m_{\kappa})
\left[\frac{2(\gamma_{1/2}+2)}{\sqrt{16\gamma_{1/2}+17}}+1\right]$ && 
$\displaystyle\frac{1}{2}\sgn(m_{\kappa})$ \\*[4ex]
3p$_{1/2}$ && 
$\displaystyle(\alpha Z)^{-2}\left[\frac{2(\gamma_{1/2}+2)}
{\sqrt{16\gamma_{1/2}+17}}-1\right]$ && 
$\displaystyle-\frac{2}{25}$ &&
$\displaystyle\frac{1}{4}\sgn(m_{\kappa})
\left[\frac{2(\gamma_{1/2}+2)}{\sqrt{16\gamma_{1/2}+17}}-1\right]$ && 
0 \\*[4ex]
3p$_{3/2}$ && 
$\displaystyle(\alpha Z)^{-2}\left[\frac{2(\gamma_{3/2}+1)}
{\sqrt{8\gamma_{3/2}+13}}-1\right]$ && 
$\displaystyle-\frac{2}{25}$ &&
$\displaystyle\frac{1}{4}\sgn(m_{\kappa})
\left[\frac{6(\gamma_{3/2}+1)}{\sqrt{8\gamma_{3/2}+13}}+1\right]$ && 
$\sgn(m_{\kappa})$ \\*[4ex]
3d$_{3/2}$ && 
$\displaystyle(\alpha Z)^{-2}\left[\frac{2(\gamma_{3/2}+1)}
{\sqrt{8\gamma_{3/2}+13}}-1\right]$ && 
$\displaystyle-\frac{2}{25}$ &&
$\displaystyle\frac{1}{4}\sgn(m_{\kappa})
\left[\frac{6(\gamma_{3/2}+1)}{\sqrt{8\gamma_{3/2}+13}}-1\right]$ && 
$\displaystyle\frac{1}{2}\sgn(m_{\kappa})$ \\*[4ex]
3d$_{5/2}$ &&
$\displaystyle(\alpha Z)^{-2}\left(\frac{2}{5}\gamma_{5/2}-1\right)$ && 
$\displaystyle-\frac{2}{25}$ && 
$\displaystyle\frac{1}{4}\sgn(m_{\kappa})
(2\gamma_{5/2}+1)$ && 
$\displaystyle\frac{3}{2}\sgn(m_{\kappa})$ \\*[4ex]
\hline
\end{tabular}
\end{center}
\end{table}
\end{landscape}
%
%\newpage
%
\begin{landscape} 
\begin{table}[h] 
\caption{Explicit forms of the coefficient
$\varepsilon_{n\kappa}^{(2)}$, defined in Eq.\ (\ref{5.3c}), for
atomic states with the principal quantum numbers $1\leqslant
n\leqslant3$. The symbol $\gamma_{\kappa}$ has been defined in Eq.\ 
(\ref{2.21}).}
\label{T.3}
\begin{center}
\begin{tabular}{lclcc}
\hline \\*[-0.5ex]
\multicolumn{1}{c}{Atomic} 
&& \multicolumn{3}{c}{$\varepsilon_{n\kappa}^{(2)}$} \\*[0.5ex]
\cline{3-5} \\*[-1.25ex]
\multicolumn{1}{c}{state} && \multicolumn{1}{c}{Exact} 
&& Nonrelativistic limit \\*[1ex]
\hline \\*[-0.5ex]
1s$_{1/2}$ && $\displaystyle\frac{1}{128}(2\gamma_{1/2}+1)
   \big(8\gamma_{1/2}^{2}+4\gamma_{1/2}-1\big)$ 
   && $\displaystyle\frac{3}{64}$ \\*[4ex]
2s$_{1/2}$ && $\displaystyle\frac{1}{128}\left[16\gamma_{1/2}^{2}
   +24\gamma_{1/2}+11+\frac{2(\gamma_{1/2}+1)
   \big(32\gamma_{1/2}^{3}+184\gamma_{1/2}^{2}
   +196\gamma_{1/2}+59\big)}{\sqrt{8\gamma_{1/2}+5}}\right]$ 
   && $\displaystyle\frac{117}{64}$ \\*[4ex]
2p$_{1/2}$ && $\displaystyle\frac{1}{128}\left[-16\gamma_{1/2}^{2}
   -24\gamma_{1/2}-11+\frac{2(\gamma_{1/2}+1)\big(32\gamma_{1/2}^{3}
   +184\gamma_{1/2}^{2}+196\gamma_{1/2}+59\big)}
   {\sqrt{8\gamma_{1/2}+5}}\right]$ 
   && $\displaystyle\frac{45}{32}$ \\*[4ex]
2p$_{3/2}$ && $\displaystyle\frac{3}{128}(2\gamma_{3/2}+1)
   \big(8\gamma_{3/2}^{2}+4\gamma_{3/2}-9\big)$ 
   && $\displaystyle\frac{45}{32}$ \\*[4ex]
3s$_{1/2}$ && $\displaystyle\frac{1}{128}\left[16\gamma_{1/2}^{2}
   +48\gamma_{1/2}+47+\frac{2(\gamma_{1/2}+2)
   \big(64\gamma_{1/2}^{3}+712\gamma_{1/2}^{2}
   +1352\gamma_{1/2}+713\big)}{\sqrt{16\gamma_{1/2}+17}}\right]$ 
   && $\displaystyle\frac{825}{64}$ \\*[4ex]
3p$_{1/2}$ && $\displaystyle\frac{1}{128}\left[-16\gamma_{1/2}^{2}
   -48\gamma_{1/2}-47+\frac{2(\gamma_{1/2}+2)
   \big(64\gamma_{1/2}^{3}+712\gamma_{1/2}^{2}
   +1352\gamma_{1/2}+713\big)}{\sqrt{16\gamma_{1/2}+17}}\right]$ 
   && $\displaystyle\frac{375}{32}$ \\*[4ex]
3p$_{3/2}$ && $\displaystyle\frac{1}{128}\left[48\gamma_{3/2}^{2}
   +72\gamma_{3/2}+9+\frac{2(\gamma_{3/2}+1)
   \big(32\gamma_{3/2}^{3}+248\gamma_{3/2}^{2}
   +356\gamma_{3/2}+75\big)}{\sqrt{8\gamma_{3/2}+13}}\right]$ 
   && $\displaystyle\frac{375}{32}$ \\*[4ex]
3d$_{3/2}$ && $\displaystyle\frac{1}{128}\left[-48\gamma_{3/2}^{2}
   -72\gamma_{3/2}-9+\frac{2(\gamma_{3/2}+1)
   \big(32\gamma_{3/2}^{3}+248\gamma_{3/2}^{2}
   +356\gamma_{3/2}+75\big)}{\sqrt{8\gamma_{3/2}+13}}\right]$ 
   && $\displaystyle\frac{525}{64}$ \\*[4ex]
3d$_{5/2}$ && $\displaystyle\frac{5}{128}(2\gamma_{5/2}+1)
   \big(8\gamma_{5/2}^{2}+4\gamma_{5/2}-25\big)$ 
   && $\displaystyle\frac{525}{64}$ \\*[4ex]
\hline
\end{tabular}
\end{center}
\end{table}
\end{landscape}
%
%\newpage
%
\begin{table}[t] 
\caption{The quantum numbers $m_{l}$ and $m_{s}$ derived from Eq.\
(\ref{5.9}) for selected values of $\kappa$ and $m_{\kappa}$.}
\label{T.4}
\begin{center}
\begin{tabular}{rcrcrcr}
\hline \\*[-0.5ex]
\multicolumn{1}{c}{$\kappa$} && \multicolumn{1}{c}{$m_{\kappa}$} && 
\multicolumn{1}{c}{$m_{l}$} && \multicolumn{1}{c}{$m_{s}$} \\*[1ex]
\hline \\*[-0.5ex]
$-\frac{1}{2}$ && $\frac{1}{2}$  && $0$  && $\frac{1}{2}$  \\*[1ex]
$-\frac{1}{2}$ && $-\frac{1}{2}$ && $0$  && $-\frac{1}{2}$ \\*[1ex]
$\frac{1}{2}$  && $\frac{1}{2}$  && $1$  && $-\frac{1}{2}$ \\*[1ex]
$\frac{1}{2}$  && $-\frac{1}{2}$ && $-1$ && $\frac{1}{2}$  \\*[1ex]
$-\frac{3}{2}$ && $\frac{3}{2}$  && $1$  && $\frac{1}{2}$  \\*[1ex]
$-\frac{3}{2}$ && $-\frac{3}{2}$ && $-1$ && $-\frac{1}{2}$ \\*[1ex]
$\frac{3}{2}$  && $\frac{3}{2}$  && $2$  && $-\frac{1}{2}$ \\*[1ex]
$\frac{3}{2}$  && $-\frac{3}{2}$ && $-2$ && $\frac{1}{2}$  \\*[1ex]
$-\frac{5}{2}$ && $\frac{5}{2}$  && $2$  && $\frac{1}{2}$  \\*[1ex]
$-\frac{5}{2}$ && $-\frac{5}{2}$ && $-2$ && $-\frac{1}{2}$ \\*[1ex]
\hline
\end{tabular}
\end{center}
\end{table}
\end{document}